\documentclass[twocolumn,pra,showpacs,floats,superscriptaddress]{revtex4}
\usepackage[dvips]{graphicx}
\usepackage{amsmath,bm}
\usepackage{psfrag}

\begin{document}

\title{Analysis of vibration impact on stability of dewetting thin liquid film}

\author{S.~Shklyaev}
\affiliation{Theoretical Physics Department, Perm State
University, Bukirev 15, Perm 614990, Russia}
\author{M.~Khenner}
\affiliation{Department of Mathematics, State University of New
York at Buffalo, Buffalo, NY 14260, USA}
\author{A.~A.~Alabuzhev}
\affiliation{Institute of Continuous Media Mechanics UB RAS, Perm
614013, Russia} \affiliation{Theoretical Physics Department, Perm
State University, Bukirev 15, Perm 614990, Russia}


\begin{abstract}
Dynamics of a thin dewetting liquid film on a vertically
oscillating substrate is considered. We assume moderate vibration
frequency and large (compared to the mean film thickness)
vibration amplitude. Using the lubrication approximation and the
averaging method, we formulate the coupled sets of equations
governing the pulsatile and the averaged fluid flows in the film,
and then derive the nonlinear amplitude equation for the averaged
film thickness. We show that there exists a window in the
frequency-amplitude domain where the parametric and shear-flow
instabilities of the pulsatile flow do not emerge. As a
consequence, in this window the averaged description is reasonable
and the amplitude equation holds.

The linear and nonlinear analyses of the amplitude equation and
the numerical computations show that such vibration stabilizes the
film against dewetting and rupture.
\end{abstract}



\date{\today}

\maketitle

\section{Introduction}

Studies of stability, dewetting and rupture of thin liquid films
on solid substrates are of great importance for micro- and
nanotechnologies. Since the first works appeared in 1960s and
1970s, these subjects continue to attract enormous attention, see
for instance Refs.~\cite{TMP,IM,GKS,PRB,WB,SHJ1,GRP,GW,BPT,LW}.
Refs.~\cite{Oron} and ~\cite{SHJ} are the reviews of recent
results in this extremely diverse field.

The subject of this paper is the general theoretical investigation
of the impact of the vibration on the stability of a thin
dewetting liquid film. Our interest in studying vibration impacts
stems from the large body of works in fluid mechanics of
macroscopic fluid layers, including
Refs.~\cite{wolf-69,wolf-70,Kozlov-98,Kozlov-01} (the experiment)
and Refs.~\cite{TVC,LChbook,Lapuerta,Thiele,Khenner} (the theory
and numerical modeling), where the vibration is shown to
drastically affect the stability characteristics and the dynamics
of fluid surfaces and interfaces.

In this paper we assume that the  external influences on the film
are the vibration, the gravity, and the  long-range molecular
attraction by the planar substrate, typified by van der Waals
forces. Other effects (such as, for instance, the
thermocapillarity and the evaporation) can be easily included by
simply adding the corresponding terms to the final evolution
equation for the film thickness.

The theory we develop is based on the standard longwave
lubrication approximation, as discussed by Oron, Davis, and
Bankoff \cite{Oron}, and on the time-averaging method.
The general discussion of the averaging methods
can be found in Refs.~\cite{S,AverBook,Nayfeh}. The key idea is
the separation of the dynamics onto fast pulsations and slow
relaxation processes. This approach works well when the vibration
frequency is high in a certain sense, i.e. when there exists a
large difference in the characteristic times (such as the viscous
relaxation time and the vibration period, see below).

The first transparent explanation of such separation of the time
scales was given by Kapitza in his pioneering work on a pendulum
with an oscillating point of support \cite{kapitza-51}. The paper
by Blekhman \cite{blekhman-00} contains other examples in
mechanics. Many examples of the successful application of the
averaging method can be found in thermal vibrational convection
\cite{TVC}, dynamics of inclusions in fluids
\cite{LChbook,Feng,bolgary}, dynamics of granular materials
\cite{evesque-rajchenbach-89}, motion of disperse fluids
\cite{Arthur}, and filtration of inclusions in porous media
\cite{Gri}.

Interestingly, the vibration of the solid plate (on which the
fluid system is located) often is capable of complete suppression
of instabilities. For example, Wolf \cite{wolf-69} experimentally
investigated the damping of the Rayleigh-Taylor instability in the
horizontal two-layer system by a vertical high frequency
vibration. The theoretical analysis of this situation (in the
linear approximation) by the averaging technique was performed by
Cherepanov \cite{AA} (summary of this paper can be found in
Ref.~\cite{LChbook}). The longwave instability in this system was
also analyzed by Lapuerta, Mancebo, and Vega \cite{Lapuerta}.
After the analysis of the linear longwave instability at the
moderate vibration frequency, they proceed to the averaged
description at high frequency. The generalization of the latter
analysis to the nonisothermal situation was developed by Thiele,
Vega, and Knobloch \cite{Thiele}. They account for the Marangoni
effect and perform a detailed investigation of the corresponding
amplitude equation.

Another widespread vibration-induced phenomenon is the parametric
excitation, which emerges when the frequency of the vibration is
comparable to one of the eigenfrequencies of the system (for
instance, to the frequency of the capillary-gravity waves).
Faraday was first to observe parametric waves on the surface of
vertically oscillating horizontal layer \cite{faraday-1831}.
Linear and nonlinear analyses of parametric instability were
performed, for instance, by Benjamin \& Ursell \cite{B&U}, Kumar
\& Tuckerman \cite{KT}, Lyubimov \& Cherepanov \cite{DV&AA}, and
by Mancebo and Vega \cite{Mancebo}. To the best of our knowledge
the latter paper is the most detailed study to-date of the linear
aspects of Faraday instability.


It must be noted that the situations termed ``the averaged motion" and
``the parametric instability" are often closely connected, although
they operate within different intervals of the vibration frequencies.
 Indeed, in the studies of the averaged
dynamics one has to ensure stability of the pulsatile motion
(periodic in time). Most fluid systems have an eigenfrequency
spectrum unbounded from above and thus the eigenfrequency is an
increasing function of the mode number. Thus, even the
high-frequency vibration is capable of parametric excitation of
the higher modes. For the pulsatile motion to be stable, a
window of parameters such as the amplitude and the frequency of the
vibration must be chosen, where the parametric instability does
not emerge.

It is also worth noting that most papers
\cite{LChbook,Thiele,TVC}, where the averaging method is employed,
deal with the vibration of ``inviscid'' frequency, i.e. the
vibration period $T_p=2\pi/\omega$ (here $\omega$ is the dimensional
frequency) is assumed small compared to the characteristic
time of viscous relaxation, $\tau_v=\hat H_0^2/\nu$ (here $\nu$ is the
kinematic viscosity and $\hat H_0$ is the mean fluid layer thickness). It is
clear that this assumption is quite reasonable for macroscopic
layers, but for thin films (of thickness $100-1000 \, \AA$) it
requires extremely large frequencies, $100 \,{\rm MHz}$ and
higher.

Now, we make a very important point, as follows.
A {\it thin film} allows for averaged description even when
{\it the viscosity is large} [i.e., $\tau_v=O\left(T_p\right)$ and even
$\tau_v\ll T_p$]: one needs only to assume that the period of the
vibration is {\it small compared to the characteristic time of
the film evolution}, $\tau_l=O\left(k^2\right)$ (here $k$ is the
wavenumber). Due to the  lubrication approximation,
$\tau_l \gg \tau_v$. Thus the condition $\tau_l \gg T_p$ is much
milder than the usual inviscid approximation $\tau_v \gg T_p$.
Therefore, the averaging procedure can be applied even to {\em
ultrathin films}.

To the best of our knowledge the only paper developing similar
approximation is Ref.~\cite{Lapuerta}, where the linear stability
problem is studied for ``moderate'' (``finite'') non-dimensional
frequencies $\omega\tau_v=O(1)$. However, by assuming the
amplitude of the vibration ``finite'' (which means that it is of
the order of the fluid layer thickness), the authors obtain that
the impact of the vibration at moderate frequency is small. Thus,
they focus on the high-frequency, small-amplitude case $\omega
\tau_v \gg 1$.

In this paper we assume {\it large} vibration amplitude
and develop the {\it nonlinear} amplitude equation for the thickness of the film.

The outline of the paper is as follows. The mathematical
formulation of the problem is presented in Sec. \ref{sec:first}.
We state and discuss assumptions and governing equations. In
particular, the separation of the processes onto fast pulsatile
and slow averaged motions is performed in
Sec.~\ref{ssec:splitting}. The pulsatile problem is solved in
Sec.~\ref{sec:puls}. The amplitude equation [Eq. (\ref{h-av_t})]
governing the averaged dynamics of the film thickness is obtained
in Sec.~\ref{sec:aver}. \textit{This equation is the main result
of this paper. It can be used to study impacts of the vertical
vibration (in the frequency range for which the averaged
description is applicable) on the dynamics of a film, in the
presence of the surface tension and wetting interactions with the
substrate.} Two limiting cases of ``low'' and ``high'' vibration
frequency are analyzed in Sec.~\ref{sec:limit}. These cases
correspond to the different ratios of the inertial and viscous forces
in the oscillatory motion. (The viscosity dominates at low
frequencies, while the inertia force dominates at high
frequencies.) In Sec. \ref{sec:3D} the 3D generalization of the
theory is presented. Conditions of parametric instability of the
oscillatory motion are analyzed in Sec.~\ref{sec:stab}, where the
Faraday instability and the shear flow instability are discussed.
We show that for any admissible vibration frequency there exists a
finite range of vibration amplitudes for which such instabilities
are not present. In Sec.~\ref{sec:layer} we address the averaged
behavior of the system within the framework of the obtained
amplitude equation. Results of linear and weakly nonlinear
analyses of the equilibrium state with the flat surface are
presented, as well as the results of direct numerical simulations.
In particular, we show that the vibration influence is
\textit{stabilizing}, i.e. it can delay or completely suppress the
film rupture by intermolecular attractive forces. Finally,
Sec.~\ref{sec:conc} summarizes the results.

\section{Formulation of the problem} \label{sec:first}

\subsection{Governing equations}

\begin{figure}[!t]
\includegraphics[width=7.0cm]{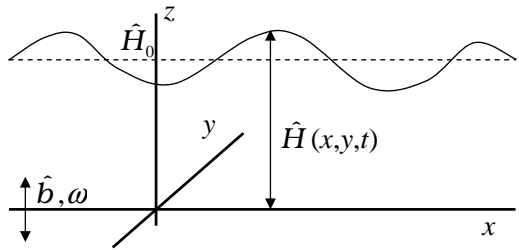}
\caption{Problem geometry.} \label{fig:1}
\end{figure}

We consider a three-dimensional (3D), laterally unbounded thin
liquid film of unperturbed thickness $\hat H_0$ on a planar, horizontal
substrate. The Cartesian reference frame is chosen such that the
$x-$ and $y-$ axes are in the substrate plane and the $z-$axis is
normal to the substrate (Fig.~\ref{fig:1}).

The substrate-film system is subjected to the vertical harmonic
vibration of the amplitude $\hat b$ and the frequency $\omega$.
Thus, in the reference frame of the substrate the acceleration of
gravity is modulated,
\begin{equation}\label{gt}
g(t)=g_0+\hat b\omega^2 \cos\omega t.
\end{equation}
Since $\hat H_0$ is small, the intermolecular interaction of the film surface with the
substrate has to be taken into account. Thus, we prescribe the
potential energy $\phi(\hat H)$ to the unit length of the film
layer. In this paper we consider only
the van der Waals attractive potential,
\begin{equation}
\phi(\hat H)=\frac{A^\prime}{6\pi \hat H^3},
\end{equation}
where $A^\prime > 0$ is the Hamaker constant \cite{Dzyalosh}. The
model presented in this paper can be readily extended to
incorporate other models of wetting interactions -- it is only
necessary to replace $\phi(\hat H)$ with an appropriate function.

We scale the time, the length, the velocity and the pressure by
$\hat H_0^2/\nu, \ \hat H_0, \ \nu/\hat H_0, \ \rho (\nu/\hat
H_0)^2$, respectively (here, $\nu$ is the kinematic viscosity and
$\rho$ is the density of the liquid). Then, the liquid motion is
governed by the following non-dimensional
problem:
\begin{subequations}
\label{base_eq}
\begin{eqnarray}
\nabla \cdot {\bf v} &=& 0,
\\
\nonumber {\bf v}_t + {\bf v}\cdot \nabla {\bf v} &=& - {\bf
\nabla} p + \nabla ^2 {\bf v} \\
&&-\left(G_0 +B \Omega^2\cos\Omega t \right) {\bf e_z},
\end{eqnarray}
\end{subequations}
\vspace{-8mm}
\begin{subequations}
\label{base_bcs}
\begin{eqnarray}
{\bf v} &=& 0 \ {\rm at} \ z = 0, \\
\nonumber h_t + {\bf v} \cdot {\bf \nabla} h &=& w,\\
 \left(p-\phi - Ca
K\right){\bf n} &=&  {\bf n}\cdot {\bf T}  \ {\rm at} \ z =
h(x,y,t).
\end{eqnarray}
\end{subequations}
Here, ${\bf v}$ is the fluid velocity, $w$ is its component normal
to the substrate, $p$ is the pressure in the liquid, $\bf T$ is
the viscous stress tensor, $h=\hat H/\hat H_0$ is the
dimensionless thickness of the layer, ${\bf e}_z$ is the unit
vector directed along the $z-$ axis, ${\bf n}= \left({\bf
e}_z-\nabla h\right)/\sqrt{1+\left(\nabla h\right)^2}$ is the
normal unit vector to the free surface, $K=\nabla \cdot {\bf n}$
is the mean curvature of the free surface, $\phi=A/h^3$ [where
$A=A'/(6\pi\rho\nu^2\hat H_0)$ is the non-dimensional Hamaker
constant], $Ca=\sigma \hat H_0/(\rho \nu^2)$ is the capillary
number (where $\sigma$ is the surface tension), $G_0=g_0 \hat
H_0^3/\nu^2$ is the Galileo number, $B=\hat b/\hat H_0$ is the
non-dimensional amplitude, and $\Omega=\omega \hat H_0^2/\nu$ is
the non-dimensional frequency.

We consider the nonlinear evolution of the large-scale
perturbations. As it is usually done, we introduce small parameter
$\epsilon$, which is of the order of the ratio of the mean
thickness $\hat H_0$ to the perturbation wavelength, i.e. $\epsilon \ll
1$ for long waves.

Below for the sake of simplicity and for more transparent
presentation of ideas, we consider the 2D model, assuming that
${\bf v}=u{\bf e}_x+ w{\bf e}_z$, where ${\bf e}_x$ is the unit
vector directed along the $x-$ axis, and all fields are
independent of $y$. The theory extends trivially in three
dimensions at the replacement of the $x$-derivative by the 2D
gradient. We derive the 3D analogue in Sec.~\ref{sec:3D}.

Next we introduce
the conventional stretched coordinates and the time:
\begin{equation}
\label{coords} X=\epsilon x, \ Z=z, \ T=\epsilon^2 t, \
\tau=\Omega t,
\end{equation}
such that $\partial/\partial t = \epsilon^2 \partial/\partial T + \Omega \partial/\partial \tau$.
Now we rescale the velocity components as follows:
\begin{equation}\label{velo}
u=\epsilon U, \ w=\epsilon^2 W.
\end{equation}
To this end, both the pressure $p$ and the surface position $h$ remain
unscaled.

Throughout this paper we assume that the capillary number is large:
\begin{equation}\label{Ca}
Ca=C\epsilon^{-2}.
\end{equation}
This is quite realistic and widely used assumption \cite{Oron}.

Substituting the expansions (\ref{coords})-(\ref{Ca}) into the problem (\ref{base_eq}) and (\ref{base_bcs}) we arrive at
the following set of equations and boundary conditions:
\begin{subequations}
\label{perturb_eq}
\begin{eqnarray}
W_Z &=& -U_X,
\\
\nonumber \Omega U_\tau &=&- p_X + U_{ZZ} \\
 && + \epsilon^2 \left(U_{XX} -
U_T - U U_X - W U_Z \right),
\\
 p_Z &=& - G_0 - B
\Omega^2\cos\tau + \epsilon^2 \left( W_{ZZ}- \Omega W_\tau\right),
\end{eqnarray}
\end{subequations}
\vspace{-8mm}
\begin{subequations}
\label{perturb_bcs}
\begin{eqnarray}
U &=& 0, \ W = 0 \ \ {\rm at} \ Z = 0, \\
\nonumber U_Z &=&- \epsilon^2 \left(W_X - 4h_X U_X\right),
\\
\nonumber \Omega
h_\tau &=& \epsilon^2 \left(W - h_T - U h_X \right),\\
\nonumber p &=& \phi - C h_{XX}  \\
&&+\epsilon^2 \left(\frac{3}{2} C h_X^2h_{XX} + 2 W_Z\right) \
{\rm at} \ Z = h.
\end{eqnarray}
\end{subequations}
All terms of order $\epsilon^4$ have been omitted.

\subsection{Separation of the pulsating and averaged dynamics}
\label{ssec:splitting}

In this paper we consider the case of moderate vibration
frequency: $\Omega \gg \epsilon^2$, i.e. the vibration period is
small compared to the characteristic time of the surface
evolution. This assumption makes possible the averaging of the
dynamics of the film over the vibration period
\cite{S,AverBook,Nayfeh}. The main purpose is the rigorous
asymptotic analysis (in powers of $\epsilon$) which results in the
set of equations and boundary conditions, where the dependence on
$\tau$ is averaged out. This set is then used to derive the
amplitude equation for the thickness of the film. As for $\Omega$,
this quantity is {\em not an asymptotic parameter}. This means
that, generally speaking, we assume $\Omega$ neither large nor
small. Therefore $\Omega$ will enter the amplitude equation as a
finite parameter.

However, within the framework of the equation set (\ref{perturb_eq}) and (\ref{perturb_bcs}) we will consider three qualitatively different cases:

(i) $\Omega \ll 1$, i.e. the vibration is of ``low'' frequency;
this is the quasi-Stokes approximation with week influence of
fluid inertia. Note that  in this
case, due to Eqs.~(\ref{coords}), the condition $\epsilon^2 \ll \Omega$ must be retained to
legalize the averaging procedure.

(ii) $\Omega=O(1)$. This assumption means that the vibration
period is comparable to
the time of the momentum  relaxation  across the layer,
$\hat H_0^2/\nu$.

(iii) $\Omega \gg 1$, i.e. the vibration is of ``high'' frequency;
viscosity is negligible except for the thin boundary layer near
the rigid wall. As usual, the boundary layer in the vicinity of
the free surface does not play an important role (see
Ref.~\cite{Mei}, for example).

Since the van der Waals interaction is important for films of
thickness $100-1000 \ \AA$, let us estimate the typical values of
$\Omega$. Taking $\hat H_0=10^{-5} \ {\rm cm}$ and $\nu=10^{-2} \
{\rm cm^2 s^{-1}}$ (viscosity of water), we obtain $\Omega \approx
10^{-4}$ for $\omega/(2\pi)=1 \ {\rm kHz}$ (typical for mechanical
vibrator), and $\Omega = 10^{-2} \div 1$ for $n=0.1 \div 10 \ {\rm
MHz}$ (typical for ultrasound). Therefore the case (i) can be
easily achieved either mechanically or by means of ultrasound
irradiation of the substrate from below, the case (ii) can be
reached only using the ultrasound, and the case (iii) seems
unrealistic. Nevertheless we shall consider this case below, since
our results, upon neglecting $\phi$, can as well be applied to the
description of macroscopic fluid layers. Besides, this limiting
case was studied in detail by many authors and therefore it allows
for the verification of our results.


We represent each field $f=\{U,W,p,h\}$ as the sum of the average
part $\bar f(T)$ and the pulsation $\tilde f(\tau)$, where $\tau$
and $T$ can now be termed the ``fast'' and the ``slow'' times,
respectively \cite{Nayfeh}. Thus, we write
\begin{subequations} \label{splitting}
\begin{eqnarray}
U &=& \bar U + \tilde U, \ W = \bar W + \tilde W,
\\
p &=& \bar p + \tilde p, \ h = \bar h + \tilde h.
\end{eqnarray}
\end{subequations}

We now assume that
\begin{equation}\label{amplit}
B=\epsilon^{-1}b,
\end{equation}
i.e. the amplitude of the vibration is {\it large} compared to the
mean film thickness. This assumption seems surprising as it is
customary to impose the {\em small-amplitude} high frequency vibration.
However, large amplitudes are permitted when the large-scale
dynamics is considered.
Indeed, it is shown below that due to
the longwave approximation the impact of the vibration becomes
non-negligible only at large amplitudes.
Also, it will be made clear momentarily that in some sense the pulsatile motion is still small-amplitude.

The assumption of large vibration amplitude means that the
oscillating part of the pressure field $\tilde p$ (which is forced by
the inertia force $B\Omega^2\cos \tau$) is of order
$\epsilon^{-1}$, which in turn leads to $\epsilon^{-1}$ scaling
for the velocities of the pulsation $\tilde U$ and $\tilde W$. The
pulsation of the surface height $\tilde h$ obviously has the same
order as $\tilde w=\epsilon^2 \tilde W$ [see Eq. (\ref{velo})],
i.e. it is of the order $\epsilon$. Therefore, it is convenient to
redefine the pulsation parts of all fields, rewriting
Eqs.~(\ref{splitting}) as follows
\begin{subequations} \label{rescaling}
\begin{eqnarray}
U &=& \bar U + \epsilon^{-1} \tilde U, \ W = \bar W +
\epsilon^{-1}\tilde W,
\\
\label{pres_scal} p &=& \bar p + \epsilon^{-1}\tilde p, \ h = \bar
h + \epsilon \tilde h,
\end{eqnarray}
\end{subequations}
where $\tilde U, \tilde W, \tilde p$ and $\tilde h$ are $O(1)$
quantities.

Accounting for the initial scaling (\ref{velo}) one can conclude
that the full components of the velocity field are:
\begin{equation} \label{rescaled_velo}
u=\epsilon \bar U + \tilde U, \ w=\epsilon^2 W + \epsilon \tilde
W,
\end{equation}
while the pressure field and the surface deviation are given by
Eqs.~(\ref{pres_scal}). Note that the pulsations of the fluid
velocity and the pressure are larger than their mean parts, while
the opposite is true for the pulsation of the film height.
Moreover,  the scaling (\ref{rescaled_velo}) means that the pulsation remains ``small-amplitude''.
Indeed, the typical horizontal (vertical) displacement of the fluid particle during one period is $O(1)$ [$O(\epsilon)$],
which is small in comparison with the respective characteristic lengthscale, $O(\epsilon^{-1})$ [$O(1)$].

Substitution of the expansions (\ref{rescaling}) in the equation
sets (\ref{perturb_eq}) and (\ref{perturb_bcs}) allows to separate
fast pulsations from background slow (averaged) motions.  Keeping
terms of zeroth and first orders in $\epsilon$, we obtain the following sets.

(i) For the pulsations:
\begin{subequations}
\label{puls_exp}
\begin{eqnarray}
\tilde W_Z &=& -\tilde U_X,
\\
\nonumber \Omega \tilde U_{\tau} &=& - \tilde p_X + \tilde U_{ZZ}
- \epsilon \left(\tilde U \tilde U_X +
\tilde W \tilde U_Z \right)_p
,
\\
\tilde p_Z &=&  - b\Omega^2\cos\tau
, \label{puls_p_z}
\\
\tilde U &=& 0, \ \tilde W = 0 \ {\rm at} \ Z = 0, \\
\nonumber \Omega \tilde h_{\tau} &=& - \tilde U \bar h_X + \tilde W- \epsilon \left[\tilde U_Z \tilde h \bar h_X
+\left(\tilde U \tilde h\right)_X\right]_p
, \\
 \tilde U_Z &=&- \epsilon \left(\tilde U_{ZZ} \tilde
h\right)_p
, \ \tilde p = -\epsilon (\tilde p_Z\tilde h)_p
\ {\rm at} \ Z = \bar h,
\end{eqnarray}
\end{subequations}
where the subscript ``$p$'' denotes the pulsating part of the
corresponding expression:
\begin{equation}
f_p=f-\bar f.
\end{equation}
(It is obvious that in general the term squared with
respect to pulsations contains both the averaged and the pulsating
components, for instance
$\cos^2\tau=\frac{1}{2}+\frac{1}{2}\cos2\tau$.)

(ii) For the averaged parts:
\begin{subequations}
\label{aver_exp}
\begin{eqnarray}
\bar W_Z &=& -\bar U_X, \  \bar p_Z = -G_0,
\\
\bar U_{ZZ} &=& \bar p_X + \langle \tilde U \tilde U_X + \tilde W
\tilde U_Z \rangle,
\\
\bar U &=& 0, \ \bar W = 0 \ {\rm at} \ Z = 0, \\
\nonumber \bar p &=&- \langle\tilde p_Z \tilde h\rangle
-\frac{\epsilon}{2}\langle\tilde p_{ZZ} \tilde h^2\rangle +
\phi(\bar h) - C \bar h_{XX}, \\
\nonumber \bar h_T &=& - \bar U \bar h_X - \langle \tilde U \tilde
h_X \rangle   + \bar W + \langle \tilde W_Z \tilde h \rangle -
\frac{\epsilon}{2}\langle \tilde U_{Z} \tilde h^2 \rangle_X,\\
\bar U_Z &=&-\langle \tilde U_{ZZ} \tilde h
\rangle-\frac{\epsilon}{2}\langle \tilde U_{ZZZ} \tilde h^2
\rangle
\label{aver_dyn}
 \ {\rm at} \ Z = \bar h.
\end{eqnarray}
\end{subequations}
In the system (\ref{aver_exp}) the angular brackets denote
averaging in time $\tau$. Note that the boundary conditions at the
free surface have been shifted at the mean position $\bar h$. This
leads to the following expansion in powers of $\epsilon$ of the
arbitrary field $F$:
\begin{equation} \label{shift}
F(Z=\bar h + \epsilon \tilde h) \approx F(\bar h) + \epsilon F_Z
(\bar h) \tilde h + \frac{1}{2}\epsilon^2 F_{ZZ} (\bar h) \tilde
h^2.
\end{equation}
As we have noted, the $\epsilon^2$-terms in Eqs.~(\ref{aver_exp})
have been omitted. On the other hand, the third term in
Eq.~(\ref{shift}) has to be taken into account for some fields,
because it produces a correction of order $\epsilon$. For
instance:
\begin{widetext}
\begin{eqnarray}
\nonumber \left< p(Z=\bar h + \epsilon \tilde h)\right> &\approx&
\left< \epsilon^{-1} \left( \tilde p+ \epsilon \tilde p_Z \tilde h
+ \frac{1}{2}\epsilon^2 \tilde p_{ZZ} \tilde h^2 \right) +\bar p +
\epsilon \bar p_Z \tilde h \right>_{Z=\bar h}\\
&=&\bar p (Z=\bar h) +\left< \tilde p_Z \tilde h +
\frac{1}{2}\epsilon \tilde p_{ZZ} \tilde h^2 \right>_{Z=\bar h}.
\end{eqnarray}
\end{widetext}
However, all terms cubic with respect to pulsations vanish after
the averaging. Moreover, in further analysis we will
disregard the $\epsilon$-terms in Eqs.~(\ref{puls_exp}) and
(\ref{aver_exp}) (i.e., use these boundary value problems in the
leading zeroth order).

It is also important to emphasize that the zeroth order of the
problem governing the pulsations [problem (\ref{puls_exp})] is
linear. This is because all nonlinear
terms ($\tilde U \tilde U_X,\ \tilde W \tilde U_Z$, etc.) are
small due to the longwave approximation, despite the scaling
(\ref{rescaled_velo}).

\section{Pulsatile motion}\label{sec:puls}

Equations.~(\ref{puls_exp}) become in the zeroth order in
$\epsilon$:
\begin{subequations}\label{puls_full}
\begin{eqnarray}
\tilde p_Z&=&-b \Omega^2 \cos \tau,\\
\Omega \tilde U_{\tau}&=&-\tilde p_X + \tilde U_{ZZ}, \ \tilde W_Z=-\tilde U_X, \\
\tilde U&=&\tilde W=0 \ {\rm at} \ Z = 0, \\
\nonumber \tilde p&=&0, \ \tilde U_Z=0, \\
\Omega \tilde h_{\tau} &=&-\tilde U \bar h_X + \tilde W \ {\rm at}
\ Z = \bar h.
\end{eqnarray}
\end{subequations}
According to the conventional method for the solution of linear
problems, we need to separate the general solution of the
homogeneous problem from the particular solution of the
nonhomogeneous one.
The
former solution corresponds to the gravity-capillary waves damped
by viscosity and thus it is not of interest. (In fact, the
gravity-capillary waves are completely damped at times of order
$\tau$. Thus the omitted solution represents fast relaxation
of the initial conditions, which has no effect on slow dynamics with
the characteristic time scale $T$.) The latter solution
corresponds to the periodically forced motion, which can be
represented in the complex form as follows:
\begin{subequations} \label{puls_complex}
\begin{eqnarray}
\tilde p &=& b \Omega {\rm Re} \left[q(X,Z)\exp{\left(i \tau\right)}\right],\\
\tilde U &=& b \Omega {\rm Re} \left[I(X,Z)\exp{\left(i\tau\right)}\right],\\
\tilde W &=& b \Omega {\rm Re} \left[K(X,Z)\exp{\left(i\tau\right)}\right],\\
\tilde h &=& b {\rm Re} \left[H(X)\exp{\left(i
\tau\right)}\right].
\end{eqnarray}
\end{subequations}
The set of equations and boundary conditions governing
the amplitudes of pulsations reads:
\begin{subequations}\label{puls}
\begin{eqnarray}
\label{puls_u} q_Z&=&-\Omega, \ I_{ZZ}+\alpha^2 I=q_X, \ K_Z=-I_X, \\
\label{puls_noslip} I&=&K=0 \ {\rm at} \ Z=0,\\
q&=&0, \ i H=K - I h_X, \ I_Z=0 \ {\rm at} \ Z=h,
\end{eqnarray}
\end{subequations}
where $\alpha^2=-i\Omega$. Hereafter we omit the bar over $\bar
h$. At $\alpha=0$ this problem coincides with the conventional
equations for the thin film in the absence of the vibration [cf.
Eqs.~(2.22)-(2.24) in Ref.~\cite{Oron} at
$\Phi=\beta_0=\Sigma=\Pi_0=\bar C^{-1}=0$]. However, the term
$\alpha^2 I$ originating from the inertia of the fluid drastically
complicates the solution.

The solution of the boundary value problem (\ref{puls}) is:
\begin{subequations}\label{sol_puls}
\begin{eqnarray}
q&=&\Omega (h-Z), \ I=i h_X\left(1-\frac{\cos\alpha (h-Z)}{\cos\alpha h}\right), \\
\nonumber K&=&-i\left[h_{XX}\left(Z+\frac{\sin\alpha
(h-Z)-\sin\alpha h}{\alpha \cos\alpha h}\right)
\right.\\
&& - \left.
 h_X^2\frac{1-\cos\alpha
Z}{\cos^2\alpha h}\right],\\
H&=&-\left[h h_{XX}f(\alpha h)-h_X^2\tan^2\alpha
h\right]
\\
&=& -\left[h f(\alpha h)h_X \right]_X,
\end{eqnarray}
\end{subequations}
where
\begin{equation} \label{f_definition}
f(y)\equiv 1-\frac{\tan y}{y}.
\end{equation}
Now, if we set $h=1+\xi$ and linearize Eqs.~(\ref{sol_puls}) with
respect to $\xi$, then we arrive at Eqs.~(2.30)-(2.32) in
Ref.~\cite{Lapuerta}. However, we have considerably expanded the
domain of validity for this solution, as will be explained in
Secs.~\ref{sec:aver} and \ref{sec:layer}.

Next, we proceed to the analysis of the limiting cases for the pulsatile
motion, i.e. $\Omega \ll 1$ (low frequency) and $\Omega \gg 1$
(high frequency).

In the former case the viscous term
$I_{ZZ}$ dominates in Eq.~(\ref{puls_u}) and the inertial term
$-i\Omega I$ exerts a week impact only. Thus the solution of the
problem (\ref{puls}) simplifies to:
\begin{subequations} \label{sol_puls_lf}
\begin{eqnarray}
q&=&\Omega (h-Z), \\
\nonumber I&=&-\frac{\Omega h_X}{2}Z(2h-Z)
\\
&&\times
\left[1-\frac{i\Omega}{12}\left(4h^2+2hZ-Z^2\right)\right], \\
\nonumber
K&=&\frac{\Omega h_{XX}}{6}Z^2\left[ 3h-Z -\frac{i\Omega
}{20}\left(20h^3-5hZ^2+Z^3\right)\right]
\\&&
+\frac{\Omega
h_X^2}{2}Z^2\left[1-i\Omega \left(h^2-\frac{Z^2}{12}\right)\right],\\
H&=&-\frac{i\Omega}{3}\left[h^3 h_X \left(1-\frac{2}{5}i\Omega
h^2\right)\right]_X.
\end{eqnarray}
\end{subequations}
Equations~(\ref{sol_puls_lf}) can also be obtained by expanding
Eqs.~(\ref{sol_puls}) in powers of the small parameter $\alpha$
and keeping all terms of order $\alpha^4$. The terms proportional
to $\Omega^2$ originate from the evolution of $\tilde U$ in
$\tau$. These terms are small corrections, but they must be
retained since they govern the solution of the averaged problem.

In the $\Omega \gg 1$ case the
amplitudes of the pulsations are [see Eqs.~(\ref{sol_puls})]:
\begin{equation} \label{high_freq_puls}
q=\Omega(h-Z), \ I=ih_X, \ K= - i h_{XX}Z, \
H=-\left(hh_X\right)_X.
\end{equation}
Equations~(\ref{high_freq_puls}) constitute the solution to the
vibration problem for an inviscid film. The longitudinal component
of the velocity, $I$, is uniform across the liquid layer; it is
determined by the pulsations of the pressure gradient only. The
transversal component, $K$, is the linear function of $Z$; it
vanishes at values of $X$ corresponding to the extremum of $I$.
Note that $H$ is real, i.e the surface deformation is in phase or
in antiphase with the vertical motion of the substrate.

Of course, $I$ given by
Eq.~(\ref{high_freq_puls}) is inconsistent with the no-slip condition
(\ref{puls_noslip}). In order to vanish $I$ at the rigid wall one
has to take the boundary layer into account. Introducing the
``fast'' coordinate $\eta=Z/\sqrt{\Omega}$ near the wall we arrive at
\begin{eqnarray}
q^{(i)}_\eta&=&0, \ iI^{(i)}-I_{\eta\eta}^{(i)}=-\Omega^{-1}q_X, \ K^{(i)}_\eta=0,\\
I^{(i)}&=&K^{(i)}=0 \ {\rm at} \ \eta=0,\\
I^{(i)} &\to& ih_X, \ q^{(i)} \to q(Z=0)=\Omega h \ {\rm at} \
\eta \to \infty.
\end{eqnarray}
The solution of this problem is well-known (see, for example,
Ref.~\cite{S}):
\begin{subequations}\label{sol-bound}
\begin{eqnarray}
q^{(i)}&=&q(Z=0)=\Omega h, \\
I^{(i)}&=&ih_X\left[1-\exp\left(-\beta \eta \right)\right], \
K^{(i)}=0,
\end{eqnarray}
\end{subequations}
where $\beta=\sqrt{i}=(1+i)/\sqrt{2}$. $|I^{(i)}(\eta)|$ increases
from zero at the rigid wall to the maximal value $1.069|h_X|$
at $\eta \approx 2.284$ and then decays to
$|I|=|h_X|$. Of course, the solution (\ref{sol-bound})
matches  the solution (\ref{sol_puls}) at $\Omega \gg 1,
\ Z \to 0$.

\begin{figure}[!t]
\includegraphics[width=8.0cm]{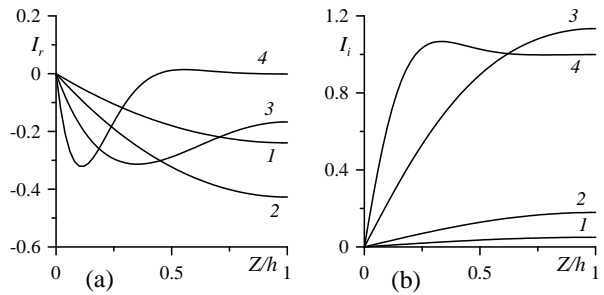}
\caption{Amplitude $I$ of the longitudinal component of the
oscillatory (pulsatile) velocity as function of $Z/h$. Real (a)
and imaginary (b) parts. Lines 1-4 correspond to $\Omega=0.5, \ 1,
\ 10, \ 100$.} \label{fig:Vz}
\end{figure}

\begin{figure}[!t]
\includegraphics[width=5.0cm]{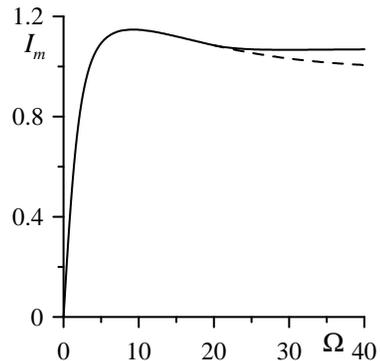}
\caption{Maximum longitudinal velocity of pulsations
$I_m=\max(|I|)$ as function of $\Omega$ (solid line); the
amplitude of this velocity at the surface (dashed line). }
\label{fig:Vm}
\end{figure}

Examples of the distribution of $I_r$ and $I_i$ across the layer
are given in Fig.~\ref{fig:Vz}(a) and Fig.~\ref{fig:Vz}(b),
respectively, for different values of $\Omega$. (Here and below we
use the subscripts ``$r$'' and ``$i$'' for the real and the
imaginary parts of complex variables.) Since $I$ is proportional
to $h_X$, the value of the latter derivative only rescales the
longitudinal velocity. Thus we set $h_X=1$ for these sketches and for Fig.~\ref{fig:Vm}.

One can see immediately that at $\Omega<\Omega_c\approx 19.74$ the
intensity of the oscillations is maximal at the free surface and
monotonously decreases to the rigid wall. At higher values of
$\Omega$ there exists a maximum in
the inner part of the layer. With increase of $\Omega$ the $Z$-coordinate
of this maximum tends to zero. At $\Omega=100$ the velocity
profile agrees well with the asymptotic formula (\ref{sol-bound}).

The dependence of the maximal amplitude of the pulsation velocity and
of $|I(Z=h)|$ on the frequency of the vibration is shown in
Fig.~\ref{fig:Vm} by the solid and dashed lines, respectively. At large $\Omega$ the solid line reaches the
asymptotical value 1.069, the dashed line tends to unity.

\begin{figure}[!t]
\includegraphics[width=6.0cm]{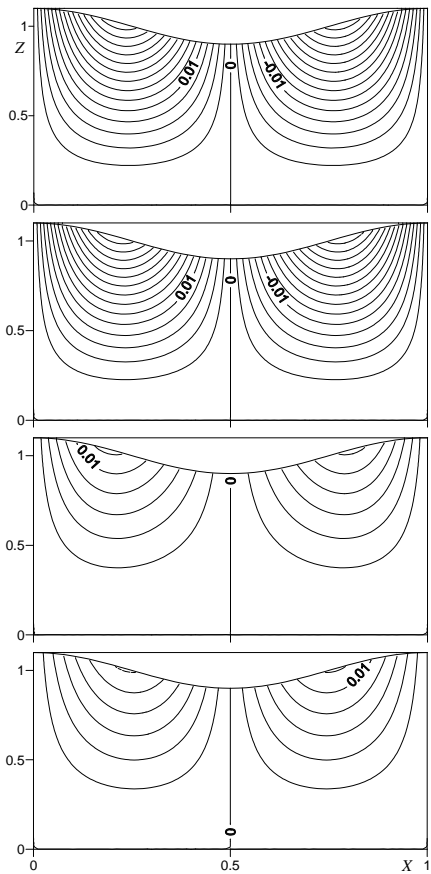}
\caption{Streamfunction $\Psi$ of pulsations for $\Omega=1$ at
$\tau=0, \pi/4, \pi/2, 3\pi/4$ (top-to-bottom).} \label{fig:Om1}
\end{figure}

\begin{figure}[!t]
\includegraphics[width=6.0cm]{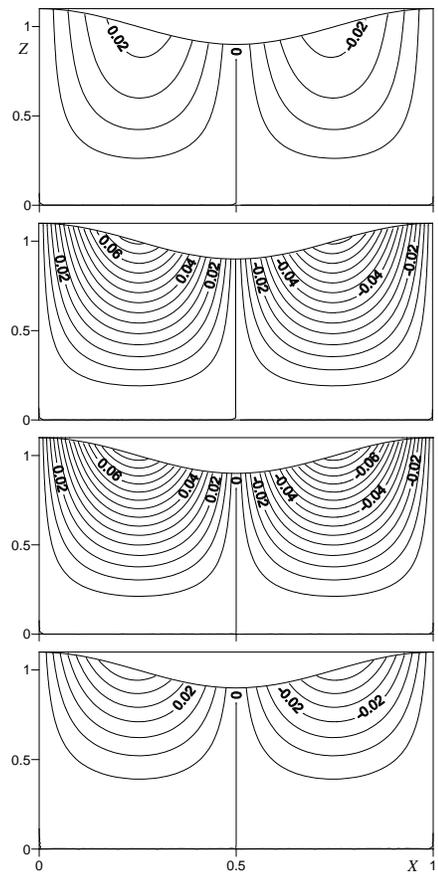}
\caption{Streamfunction $\Psi$ of pulsations for $\Omega=10$ at
$\tau=0, \pi/4, \pi/2, 3\pi/4$ (top-to-bottom).} \label{fig:Om10}
\end{figure}

To make more clear the behavior of the pulsation velocity we show
in Figs.~\ref{fig:Om1} and ~\ref{fig:Om10} the isolines of the
pulsation streamfunction at four progressive time moments. The
streamfunction $\Psi$ is defined as
\begin{equation}
\Psi=i h_X\left[Z+\frac{\sin\alpha (h-Z)-\sin\alpha h}{\alpha
\cos\alpha h}\right],
\end{equation}
so that $I=\Psi_Z, \ K=-\Psi_X$. Again for illustration purposes only we set $h=1+a \cos
kX$ with $a=0.1, \ k=1$ in these figures.

We point out that these figures present the streamfunctions of the
time-periodic motion, i.e. one must not be concerned that the
isolines are open curves. Each fluid particle oscillates near its
mean position with the small (on the scale of the figure) amplitude,
and its instantaneous pulsation velocity is tangential to the
momentary isoline at the point.


It must be emphasized that within the framework of the longwave
approximation there is no need for the stability analysis of the
solution (\ref{sol_puls}). Indeed, the leading order of
Eqs.~(\ref{puls_exp}) [given by Eqs.~(\ref{puls_full})] is the
linear problem. As was mentioned above, the homogeneous problem
originating from Eqs.~(\ref{puls_full}) obviously has only the
decaying solutions: at finite $\Omega$, the perturbations decay
due to the viscosity. Thus, stability of the oscillatory motion is
evident except for the limiting case $\Omega \gg 1$, which has
been studied in detail \cite{B&U,Mancebo,DV&AA,LChbook}. However,
stability with respect to perturbations of finite or small
wavelength must be addressed. We briefly discuss this issue in
Sec.~\ref{sec:stab}. Also, in the Appendix \ref{app:p-flow}
we discuss the reduction of the flow (\ref{sol_puls}) to the
oscillatory Poiseuille flow. The analysis in Sec.~\ref{sec:stab}
is partially based on this result.

\section{Averaged motion} \label{sec:aver}

In this section we solve the leading order of the averaged problem (\ref{aver_exp}):
\begin{subequations}\label{aver0}
\begin{eqnarray}
\bar p_Z&=&-G_0, \ \bar W_Z=-\bar U_X, \\
\bar U_{ZZ}&=&\bar p_X+\langle{\tilde U \tilde U_X + \tilde W \tilde U_Z}\rangle, \\
\bar U&=&\bar W=0 \ {\rm at} \ Z = 0, \\
\nonumber \bar p&=&\phi( h)-C h_{XX} + b \Omega^2 \langle{\tilde h
\cos \tau}\rangle, \\
\nonumber h_T &=&  \bar W -\bar U h_X+ \langle{\tilde h \tilde W_Z
- \tilde
U \tilde h_X }\rangle, \\
\bar U_Z&=&-\langle \tilde U_{ZZ} \tilde h\rangle \ {\rm at} \ Z =
h.
\end{eqnarray}
\end{subequations}
Note that the averaged term in the boundary condition for the
pressure results when $\tilde p_Z$ in
the corresponding boundary condition (\ref{aver_dyn}) is replaced
by $-b\Omega^2 \cos \tau$ [see Eqs.~(\ref{puls_full})].

Substituting the forms (\ref{puls_complex}), accounting for the
obvious equality
\begin{equation} \label{aver_LL}
\langle {{\rm Re} \left( B e^{i \tau} \right){\rm Re} \left( D
e^{i \tau} \right)}\rangle=\frac{1}{2}{\rm Re} \left(B D^*\right)
\end{equation}
for the calculation of averages in Eqs.~(\ref{aver0}), and noting that
\begin{equation}
\langle \tilde U\tilde h_X - \tilde W_Z \tilde h\rangle= \langle
\tilde U \tilde h_X + \tilde U_X \tilde h\rangle = \langle \tilde
U \tilde h\rangle_X,
\end{equation}
we obtain
(the overbars are omitted):
\begin{subequations}\label{aver}
\begin{eqnarray}
\label{aver_velo}
p_Z&=&-G_0, \ W_Z=-U_X, \\
U_{ZZ}&=&p_X+\frac{1}{2} b^2 \Omega^2{\rm Re} \left(I^* I_X+
K^*I_Z\right),\\
U&=&W=0 \ {\rm at} \ Z = 0, \\
\nonumber p&=&\phi-C h_{XX} + \frac{1}{2} b^2 \Omega^2{\rm Re} H, \\
\nonumber h_T &=&- U h_X +W - \frac{1}{2}b^2 \Omega {\rm Re}
\left(I^*
H\right)_X, \\
U_Z&=&-\frac{1}{2}b^2 \Omega {\rm Re} \left(I_{ZZ}^* H\right) \
{\rm at} \ Z = h.
\end{eqnarray}
\end{subequations}
The evolutionary equation for $h$ can be rewritten in
the form
\begin{equation} \label{int_aver}
h_T =- \partial_X \int_0^h U dZ - \frac{1}{2} b^2 \Omega {\rm Re}
\left(I^* H\right)_X \ {\rm at} \ Z = h.
\end{equation}

\begin{figure}[!t]
\includegraphics[width=8.0cm]{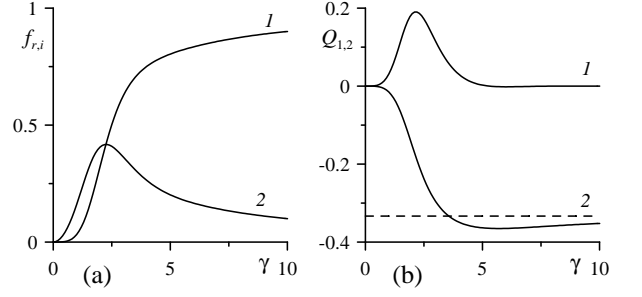}
\caption{Coefficients in the amplitude equation (\ref{h-av_t}) as
functions of $\gamma=\sqrt{2\Omega}h$: (a): Real and imaginary
part of $f(\alpha h)$, (b): $Q_{1,2}(\gamma)$. } \label{fig:Qf}
\end{figure}

Analytical integration of this set of equations is performed in
Appendix~\ref{app-aver}. It results in the following nonlinear
equation for $h$:
\begin{subequations}\label{h-av_t}
\begin{eqnarray}
h_T&=&\left(\frac{1}{3}h^3 \Pi_X-\frac{1}{2}b^2 \Omega^2Q\right)_X, \\
\nonumber \Pi & \equiv& p(Z=0)\\
&=& \phi(h)-C h_{XX} + G_0 h
+\frac{1}{2}b^2 \Omega^2{\rm Re} H.\\
Q&=&Q_1(\gamma)h^2h_X^3+Q_2(\gamma)h^3h_Xh_{XX},
\end{eqnarray}
\end{subequations}
where $\gamma=\sqrt{2\Omega}h$ and
\begin{subequations}\label{coeff}
\begin{eqnarray}
Q_1&=&3\frac{2\sinh\gamma \sin\gamma-\gamma\left(\sinh\gamma\cos\gamma+\sin\gamma\cosh\gamma\right)}
{\gamma^2\left(\cos\gamma+\cosh\gamma\right)^2}, \\
Q_2&=&-\frac{1}{3}+\frac{11\left(\sinh\gamma-\sin\gamma\right)-
3\gamma\left(\cosh\gamma-\cos\gamma\right)}{\gamma^3\left(\cos\gamma+\cosh\gamma\right)}.
\end{eqnarray}
\end{subequations}
Also note that the function $f(\alpha h)$ [see Eq.~(\ref{f_definition})] can be expressed
in terms of $\gamma$:
\begin{subequations} \label{f_gam}
\begin{eqnarray}
f_r(\alpha
h)&=&1-\frac{\sinh\gamma+\sin\gamma}{\gamma\left(\cos\gamma+\cosh\gamma\right)},\\
f_i(\alpha h)
&=&\frac{\sinh\gamma-\sin\gamma}{\gamma\left(\cos\gamma+\cosh\gamma\right)}.
\end{eqnarray}
\end{subequations}
The dependence of functions $f_{r,i}(\alpha h)$ and $Q_{1,2}$ on
the parameter $\gamma$ is given in Figs.~\ref{fig:Qf}(a) and \ref{fig:Qf}(b), respectively. One can see that
$0\le f_r(\alpha h)<1$ and the coefficient $Q_2$ is negative for
all values of $\gamma$.

One can immediately see that along with the regular
contributions due to the wetting potential, surface tension and gravity, the
expression for $\Pi$ contains the
nonlinear contribution due to the vibration. The nonlinear $Q$-term is
entirely due to the vibration.

The obtained equations of slow motion allow simplification in the
limits of low and high frequency $\Omega$. For these limiting
cases, considered next, the solution to the boundary value problem
(\ref{aver}) is not very cumbersome and can be presented in
detail.

\section{Frequency-based analysis of Eqs. (\ref{aver}) and (\ref{h-av_t})} \label{sec:limit}

\subsection{Low vibration frequency, $\Omega \ll 1$}

We look for the solution of the averaged
problem (\ref{aver}) in the form of the power series in $\Omega$,
using the corresponding solution (\ref{sol_puls_lf}) of the
pulsatile problem.

It is clear that $W$ can be easily expressed via $U$ by means of
the continuity equation, the last relation in
Eqs.~(\ref{aver_velo}). However, the corresponding expression
involves different combinations of derivatives of $h$ and is
difficult to understand. Thus, for the sake of
brevity, here we present only the longitudinal component of the
averaged velocity:
%
\begin{subequations} \label{sols_lowfreq}
\begin{eqnarray}
\nonumber U&=&\frac{1}{2}Z(Z-2h)\Pi_X+\frac{\left(b \Omega^2\right)^2}{720}Z\left[75h_X^3 hZ^3\right.  \\
&& \left. + h_X
h_{XX}\left(Z^5-6hZ^4+15h^2Z^3-48h^5\right)\right], \label{Usols_lowfreq}\\
 \Pi&=&\phi(h)-C h_{XX} +G_0 h -
\frac{\left(b\Omega^2\right)^2}{15}(h^5 h_X)_X.
\end{eqnarray}
\end{subequations}
(Also note that the pressure $p=\Pi-G_0 Z$.)
The terms proportional to $\Omega^4$ are retained in this
solution, while the higher order terms have been omitted. The
first term in Eq.~(\ref{Usols_lowfreq}) is conventional, while the
second one represents the impact of the vibration (as well as the
last term in $\Pi$).

Substituting Eqs. (\ref{sols_lowfreq}) in Eq.~(\ref{int_aver}), we obtain
the evolutionary equation for $h$:
\begin{equation} \label{h_aver_low}
h_T =\frac{1}{3}\left(h^3\Pi_X\right)_X + \left(b
\Omega^2\right)^2 \left(\frac{2}{63}h^7h_Xh_{XX}-\frac{1}{15}h^6
h_X^3\right)_X.
\end{equation}
Equation~(\ref{h_aver_low}) can be obtained also from
Eq.~(\ref{h-av_t}) by noticing that at small $\gamma$ the
coefficients $Q_{1,2}$ (and $f_r$) are proportional to $\gamma^4$:
\begin{eqnarray} \label{small}
f_r(\alpha h)\approx& Q_1\approx \frac{1}{30} \gamma^4, \ Q_2
\approx -\frac{1}{63} \gamma^4.
\end{eqnarray}
Equation~(\ref{h_aver_low}) makes it clear that it is necessary to
provide large vibration amplitude in order to gain the finite
impact of the vibration on the dynamics of the film height. In
other words, the rescaled {\it acceleration} $b \Omega^2$ has to
be finite. Moreover, we show in Sec.~\ref{sec:stab} that such
values of the amplitude do not cause the parametric instability.


It is important to recognize that one needs only
$\gamma \ll 1$ in order to obtain Eq.~(\ref{h_aver_low}). However,
this limit can be reached not only for small $\Omega$, but
also for small {\em local} values of $h$. This means that
Eq.~(\ref{h_aver_low}) can be used (independently of the value of
$\Omega$) near rupture, i.e. in the close vicinity of the point $X_r$,
such that $h(X_r,T)\to 0$. Note that the general
Eq.~(\ref{h-av_t}) should be applied far away from this point.

However, since the vibration terms in Eq.~(\ref{h_aver_low}) are
proportional to $h^9$, they are negligible at small $h$ in
comparison with the term originating from the surface tension
(which is of order $h^4$) and with the dominant van der Waals
term. In other words, only the competition of the surface tension
and the van der Waals interaction governs the behavior of the film
near rupture, and the vibration does not provide a noticeable
impact. (Of course, in the very close vicinity of the rupture
point only the van der Waals interaction contributes to the film
dynamics -- see Refs.~\cite{Oron}.)

\subsection{High vibration frequency, $\Omega \gg 1$}

Here we
use the `inviscid'' solution for the pulsations given by
Eqs.~(\ref{high_freq_puls}).
We do not consider the impact of the boundary layer, i.e. the
solution given by Eq.~(\ref{sol-bound}), for the obvious reason.
It is known from Rayleigh \cite{Rayleigh} and Schlichting
\cite{S}, that the boundary layer can produce an independent
averaged motion. However, the intensity of this flow is rather
small in comparison with the volumetric sources under
consideration. Indeed, estimating the longitudinal component of
the averaged velocity generated at the external border of the
boundary layer, one obtains $U_S \propto b^2 \Omega
\left(|I|^2\right)_X$, whereas the dominant contribution in  the
Eq.~(\ref{aver_velo}) is proportional to $b^2 \Omega^2
\left(|I|^2\right)_X$. A similar situation exists in many problems
of thermal vibrational convection \cite{TVC}.

Now one can see
that in the set (\ref{aver})
\begin{subequations} \label{hf_mean}
\begin{eqnarray}
{\rm Re}H&=&H=-\left(h h_X\right)_X, \\
{\rm Re} \left(I^* I_X+ K^*I_Z\right)&=&h_X
h_{XX}=\frac{1}{2}\left(h_X^2\right)_X.
\end{eqnarray}
\end{subequations}
The corresponding terms are proportional to $\Omega^2$. Other
averaged terms in Eqs.~(\ref{aver}) are proportional to $\Omega$,
and thus they can be safely neglected.

The calculation gives
\begin{subequations} \label{u-hifr}
\begin{eqnarray}
p&=&\Pi-GZ, \\
\Pi&=&\phi(h)-C h_{XX} +G_0 h -
\frac{b^2\Omega^2}{2}(h h_X)_X, \\
U&=&\frac{1}{2}Z(Z-2h)\left(\Pi+\frac{b^2 \Omega^2}{4}
h_X^2\right)_X,
\end{eqnarray}
\end{subequations}
which in view of Eq.~(\ref{int_aver}) results in the following equation governing the evolution of the
thin film thickness:
\begin{eqnarray} \label{h_t-hf}
\nonumber h_T&=&\frac{1}{3}\left\{h^3 \left[\phi(h)-C h_{XX} + G_0
h
\right.\right.\\
&& - \left.\left. \frac{b^2\Omega^2}{4}
\left(2hh_{XX}+h_X^2\right)\right]_X\right\}_X.
\end{eqnarray}
Since for $\gamma \gg 1$
\begin{eqnarray} \label{large}
f_r(\alpha h)&\approx 1-\gamma^{-1}, \ Q_1 = e.s.t., \ Q_2 \approx
-\frac{1}{3} -\frac{3}{\gamma^2},
\end{eqnarray}
Eq.~(\ref{h_t-hf}) also follows directly from
Eq.~(\ref{h-av_t}) in the high frequency approximation. Here
``e.s.t.'' denotes the exponentially small term.

Obviously, at finite $b$ and large $\Omega$ the vibration
determines the film dynamics at large. One has to assume that $b
\ll 1$ to retain the competition of the surface tension and
gravity in the evolution of the film. Such small amplitude, high
frequency vibration has been subject of many papers
\cite{AA,Lapuerta,Thiele}. Equation~(\ref{h_t-hf}) coincides with
the equation derived, in the high frequency approximation, by
Lapuerta et al. \cite{Lapuerta} (see also Ref.~\cite{Thiele}).
Formally, we obtain that the same equation remains valid at larger
amplitudes, but in fact finite values of $b$ cannot be reached for
$\Omega \gg 1$ due to the instability of the pulsatile motion (see
Ref.~\cite{Mancebo} and Sec.~\ref{sec:stab}).

Finally, we recall from Sec.~\ref{ssec:splitting} that the high
frequency limit is equivalent to the approximation of thick fluid
layer. Thus the  Eq.~(\ref{h_t-hf}), strictly speaking, can be
applied only with $\phi=0$.

\section{3D case} \label{sec:3D}

In this section we generalize the theory to the 3D
case. The starting point is the 3D analogue of
Eqs.~(\ref{perturb_eq}) and (\ref{perturb_bcs}):
\begin{subequations}
\label{perturb_eq_3D}
\begin{eqnarray}
W_Z &=& -{\bf \nabla} \cdot {\bf U},
\\
\nonumber \Omega {\bf U}_\tau  &=&  -{\bf \nabla}p + {\bf U}_{ZZ} + \epsilon^2 \left(\nabla^2{\bf U} - {\bf U}_T\right) \\
&& -\epsilon^2 \left({\bf U} \cdot {\bf \nabla} {\bf U} + W {\bf
U}_Z \right),
\\
 p_Z &=&-  G_0 - B
\Omega^2\cos\tau + \epsilon^2 \left( W_{ZZ}- \Omega W_\tau\right),
\end{eqnarray}
\end{subequations}
\vspace{-8mm}
\begin{subequations}
\label{perturb_bcs_3D}
\begin{eqnarray}
{\bf U} &=& 0, \ W = 0 \ \ {\rm at} \ Z = 0, \\
\nonumber {\bf U}_Z &=&- \epsilon^2 \left[{\bf \nabla} W -
2 {\bf \nabla}h {\bf \nabla} \cdot {\bf U} - {\bf \nabla}h \cdot \left({\bf \nabla U}+{\bf \nabla U}^T\right)\right],\\
\nonumber \Omega
h_\tau &=& \epsilon^2 \left(W - h_T - {\bf U}\cdot {\bf \nabla} h \right),\\
\nonumber p &=& \phi - C \nabla^2 h
\\
&& +\epsilon^2 \left[\frac{3}{2} C \left(\nabla h\right)^2
\nabla^2 h + 2 W_Z\right] \ {\rm at} \ Z = h.
\end{eqnarray}
\end{subequations}
Here $\nabla=\left(\partial_X,
\partial_Y, 0\right)$ is the 2D gradient, ${\bf U}$ is the
projection of the velocity onto the $X-Y$ plane, i.e. ${\bf v} =
\epsilon {\bf U} + \epsilon^2 W {\bf e_z}$, $\left({\bf \nabla
U}\right)$ is second-order tensor [i.e., $\left({\bf \nabla
U}\right)_{jl}=\partial U_l/\partial x_j,\ j,l=1,2$], and other
notations are unchanged.

It is easy to see that in the leading order ($\epsilon^0$)
the only difference between the systems (\ref{perturb_eq}) and
(\ref{perturb_bcs}) (2D case) and (\ref{perturb_eq_3D}) and
(\ref{perturb_bcs_3D}) (3D case) is the replacement of $U$ by
${\bf U}$ and $\partial_X$ by ${\bf \nabla}$. Less evident is that
only same changes are warranted in the solution as well. This can
be easily checked by repeating the analysis quite similar to the
one presented in Secs.~\ref{sec:first}-\ref{sec:limit}. Here we
show the results only.

The solution of the problem for the pulsations is  [cf.
Eqs.~(\ref{sol_puls})]:
\begin{subequations}\label{sol_puls_3D}
\begin{eqnarray}
q&=&\Omega (h-Z), \ {\bf I}=i \left[1-\frac{\cos\alpha (h-Z)}{\cos\alpha h}\right]{\bf \nabla}h, \\
\nonumber K&=&-i\left[Z+\frac{\sin\alpha (h-Z)-\sin\alpha
h}{\alpha \cos\alpha h}\right]\nabla^2 h
\\
&& +i\frac{1-\cos\alpha
Z}{\cos^2\alpha h}\left({\bf \nabla}h\right)^2,\\
H&=&-{\bf \nabla} \cdot \left[h f(\alpha h){\bf \nabla}h \right],
\end{eqnarray}
\end{subequations}
while the averaged dynamics of the free surface is governed by the
following equation:
\begin{subequations}\label{h-av_t_3D}
\begin{eqnarray}
h_T&=&{\bf \nabla}\cdot \left[\frac{1}{3}h^3 {\bf \nabla} \Pi-\frac{b^2\Omega^2}{2}{\bf Q}(h)\right], \\
\nonumber \Pi& \equiv& p(Z=0)\\
&=& \phi(h)-C \nabla^2h + G_0 h
+\frac{b^2\Omega^2}{2}{\rm Re} H.\\
{\bf Q}&=&Q_1(\gamma)h^2\left(\nabla h\right)^2{\bf \nabla}h
+Q_2(\gamma)h^3 \nabla^2 h {\bf \nabla}h,
\end{eqnarray}
\end{subequations}
with $Q_{1,2}$ and $f$ given by Eqs.~(\ref{coeff})-(\ref{f_gam}),
respectively.

In the limiting cases Eq.~(\ref{h-av_t_3D})  simplifies as
follows:

(i) Low frequency, $\Omega\ll 1$
\begin{eqnarray}
\label{h_aver_low_3D} \nonumber h_T &=&\frac{1}{3}{\bf
\nabla}\cdot\left(h^3\ {\bf \nabla}\Pi\right)
\\
&& + \left(b \Omega^2\right)^2 {\bf \nabla}\cdot
\left[\frac{2h^7}{63} \nabla^2h {\bf \nabla}h-\frac{h^6}{15}
\left(\nabla h\right)^2 {\bf \nabla}h\right],\\
 \Pi&=&\phi(h)-C \nabla^2h +G_0 h -
\frac{b^2\Omega^4}{15}{\bf \nabla}\cdot (h^5 {\bf \nabla}h).
\end{eqnarray}

(ii) High frequency, $\Omega\gg 1$
\begin{eqnarray}
\label{h_t-hf_3D} h_T&=&\frac{1}{3}{\bf \nabla}\cdot \left\{h^3
{\bf \nabla}\left[\Pi+ \frac{b^2\Omega^2}{4}\left({\bf
\nabla}h\right)^2\right]\right\},
\\
 \Pi&=&\phi(h)-C \nabla^2h +G_0 h -
\frac{b^2\Omega^2}{2}{\bf \nabla}\cdot\left(h {\bf
\nabla}h\right).
\end{eqnarray}
Omitting the van der Waals interaction potential $\phi(h)$ one
immediately reduces the last equation to Eqs.~(3.11) and (3.12) in
Ref.~\cite{Lapuerta} [see also Eq.~(3.19) in Ref.~\cite{Thiele} at
$Ma=0$].

\section{Stability of the pulsatile flow} \label{sec:stab}
\subsection{Analysis of the deformable mode}\label{ssec:finitek}

In this section we analyze stability of the pulsatile motion given
by Eqs.~(\ref{sol_puls}). It was mentioned already in
Sec.~\ref{sec:puls} that there is no any doubt about its stability
within the framework of the longwave approximation,
Eqs.~(\ref{puls_full}). However, the question whether this motion
is stable with respect to perturbations with {\em shorter
wavelength}  is quite reasonable, especially in view of unusual
scaling (\ref{amplit}): the amplitude of the vibration is
large (of order $\epsilon^{-1}$) and one can expect the emergence
of the parametric instability.

Therefore, we return to the governing
equations (\ref{base_eq}) and  (\ref{base_bcs}) and study the
stability of the ``base state'':
\begin{eqnarray} \label{stab_basic}
p_0&=&\epsilon^{-1}b\Omega{\rm Re} \left[ q(\epsilon
x,z)\exp\left(i\Omega t\right)\right] +
O(1),\\
u_0&=&b\Omega{\rm Re} \left[I(\epsilon x,z)\exp\left(i\Omega
t\right)\right] +
O(\epsilon),\\
w_0&=&O(\epsilon), h_0=h_0(\epsilon x,\epsilon^2 t)+O(\epsilon).
\end{eqnarray}
[See Eqs.~(\ref{pres_scal}) and (\ref{rescaled_velo}) for scalings
and Eqs.~(\ref{puls_complex}) for the pulsation velocity and
pressure.] We restrict analysis to the 2D {\em base state} since
the stability problem for the 3D base state [$h_0=h_0(\epsilon x,
\epsilon y,\epsilon^2 t)$] admits reduction to the one with the 2D
base state. This will be shown below. Introducing small
perturbations and linearizing the problem near the base state we
obtain:
\begin{subequations}
\label{pert_eq}
\begin{eqnarray}
u_x + w_z &=& 0,
\\
u_t + u_0 u_x + w u_{0z}  &=& - p_x + \nabla ^2 u,
\\
w_t + u_0 w_x &=& - p_z + \nabla ^2 w,
\end{eqnarray}
\end{subequations}
\vspace{-8mm}
\begin{subequations}
\label{pert_bcs}
\begin{eqnarray}
u &=& 0, \ w = 0 \ \ {\rm at} \ z = 0, \\
\nonumber \xi_t&=& w -u_0 \xi_x,\ \ u_z + w_x =-u_{0zz}\xi,\\
\nonumber p &=& \left[\epsilon^{-1} b\Omega^2 \cos\Omega t + G_0+
\phi'(h_0)\right]\xi \\
&&- \epsilon^{-2}C \xi_{xx} - 2w_z =0 \ {\rm
at} \ z = h_0,
\end{eqnarray}
\end{subequations}
where $\xi$ is the perturbation to the surface deflection, and the
obvious notations $\{p,u,w\}$ are used for other
perturbations. Again, we will show below that the 3D {\it
perturbations} do not warrant the
consideration.

Note that for this analysis one can safely neglect variation of
$h_0$ on the time scale $T=\epsilon^2t$ and on the length scale
$X=\epsilon x$. Thus the unperturbed surface is assumed to be
locally flat. Same approximation is also appropriate for the
velocity components, thus we consider the plane-parallel flow with
vanishing transversal component $w_0$ and the longitudinal
component nearly constant in $x$. The obtained problem is quite
similar to the problem governing the Faraday instability (see, for
example, Ref.~\cite{Mancebo}). The only difference is the presence
of the base flow $u_0$ in Eqs.~(\ref{pert_eq}) and
(\ref{pert_bcs}).

Presenting all fields in the form of normal perturbations
$\left(u,w,p,\xi\right)=\left(\hat u(t,z),\hat w(t,z),\hat
p(t,z),\hat \xi(t,z)\right)e^{ikx}$, where $k$ is the wavenumber
of the perturbation, we obtain the following problem for the
amplitudes:
\begin{subequations}
\label{pert_eq_k}
\begin{eqnarray}
ik \hat u + \hat w_z &=& 0,
\\
\hat u_t + ik u_0 \hat u + \hat w u_{0z}  &=& - ik\hat p + \hat u_{zz}-k^2\hat u,
\\
\hat w_t + ik u_0 \hat w &=& - \hat p_z + \hat w_{zz}-k^2\hat w ,
\end{eqnarray}

\end{subequations}
\vspace{-8mm}
\begin{subequations}
\label{pert_bcs_k}
\begin{eqnarray}
\hat u &=& 0, \ \hat w = 0 \ \ {\rm at} \ z = 0, \\
\nonumber \hat \xi_t&=& \hat w - ik u_0 \hat \xi,\ \ \hat u_z + ik \hat w =-u_{0zz}\hat \xi,\\
\nonumber \hat p &=& \left[\epsilon^{-1} b\Omega^2 \cos\Omega t +
G_0+
\phi'(h_0)\right]\hat \xi \\
&&+ \epsilon^{-2}Ck^2 \hat \xi - 2\hat w_z =0 \ {\rm at} \ z =
h_0.
\end{eqnarray}
\end{subequations}

This problem contains terms of different orders with respect
to small parameter $\epsilon$, which simplifies the analysis.
Indeed, the term $\epsilon^{-2}C \xi_{xx}$ prevails in the
boundary condition and produces the stabilizing effect, since it
suppresses deviations of the surface. Instability may occur when
this term is comparable to the potentially destabilizing term
$\epsilon^{-1} b\Omega^2 \cos\Omega t$ in the same boundary
condition. This is only possible for long waves with
$k=\sqrt{\epsilon}K$. It can be shown easily that for other $k$
the perturbations decay, except for the case considered in
Sec.~\ref{ssec:nondef}.

Choosing the following scalings for the perturbations (see
Ref.~\cite{Mancebo}):
\begin{eqnarray} \label{pert_sca}
\hat p=\hat P, \ \hat u = \sqrt{\epsilon} \hat U, \ \hat w=\epsilon
\hat W, \ \hat \xi=\epsilon \hat \Xi,
\end{eqnarray}
we obtain in the leading order
\begin{subequations} \label{pert_Mancebo}
\begin{eqnarray}
\hat U_t &=&- iK\hat P + \hat U_{zz}, \ iK \hat U + \hat W_z = 0, \
\hat P_z  = 0,\\
\hat u &=& 0, \ \hat w = 0 \ \ {\rm at} \ z = 0, \\
\nonumber \hat \Xi_t&=& \hat W, \ \hat U_z =0,\\
\hat P &=& b\Omega^2 \cos\Omega t \hat \Xi + C K^2 \hat \Xi  =0 \
{\rm at} \ z = h_0.
\end{eqnarray}
\end{subequations}
It can be seen that in Eqs.~(\ref{pert_eq_k}) and
(\ref{pert_bcs_k}) the terms containing base flow $u_0$ are of low
order, and thus they dropped out of Eqs.~(\ref{pert_Mancebo}).
This means that within the framework of scaling (\ref{pert_sca})
Faraday instability prevails over the instability due to shear
flow  (see Sec.~\ref{ssec:nondef} for the opposite case).

Next, it is evident that there is no preferential direction in the
$x-y$ plane for the problem (\ref{pert_Mancebo}). This allows the
reduction of the stability with respect to 3D perturbations to 2D
problem under consideration: one needs only to choose the $x$-axis
in the direction of perturbations' wavevector. Moreover, since the
velocity of the base state also dropped out from the leading order
of the stability problem, the symmetry properties of the base
state are inessential. Therefore, even for 3D base state the
stability is governed by the same Eqs.~(\ref{pert_Mancebo}).

Thus, we showed that the problem under consideration completely
reduces to the analysis of Faraday instability. Such analysis was
performed in detail by Mancebo and Vega \cite{Mancebo}.
Equations~(\ref{pert_Mancebo}) can be rewritten in the form of
Eqs.~(2.11)-(2.13) in their paper. Indeed, the case considered
here corresponds to the case B.1.2 (Long-wave limit) in their
paper. Introducing the parameters as in Ref.~\cite{Mancebo}, we
obtain:
\begin{eqnarray} \label{par_Mancebo}
\tilde \omega_{MV}=\Omega h_0^2, \ \tilde a_{MV}=\frac{b}{\sqrt{C}h_0^{3/2}}, \ \gamma_{MV} = \epsilon \frac{G_0}{\sqrt{C}}h_0^{5/2} \ll 1.
\end{eqnarray}
Here the subscript ``MV'' is used to mark the parameters used by
Mancebo and Vega \cite{Mancebo}. Note that these parameters must
be calculated using the local thickness $\hat H=\hat H_0 h_0$
instead of the mean value $\hat H_0$, which causes the appearance
of $h_0$ in Eqs.~(\ref{par_Mancebo}).

The critical value of the acceleration $\tilde a_{MV,c} \tilde
\omega_{MV}^2$ as a function of $\tilde \omega_{MV}$ is presented
in Fig.~4 of Ref.~\cite{Mancebo}. Thus, the critical value of the
amplitude is
\begin{equation}\label{b_c}
b_c=\frac{\sqrt{C}\Phi_{MV}(\Omega h_0^2)}{\Omega^2 h_0^{5/2}},
\end{equation}
where $\Phi(\mu)$ is the function given in Ref.~\cite{Mancebo}
(Fig.~4 there). Due to Eq.~(\ref{par_Mancebo}) we are interested
only in the line corresponding to $\gamma_{MV}=0$.

At $\mu \to 0$, $\Phi_{MV}(\mu)\approx 8.5$, then it decreases to approximately $5.5$ at
$\mu \approx 6$ and after that grows. At large $\tilde \omega_{MV}$ the asymptotic formula
\begin{equation}
\Phi_{MV}(\mu) \approx \sqrt{\mu}
\end{equation}
holds. It means that
\begin{equation}
b_c\left(\Omega h_0\right)^{3/2}C^{-1/2} \to 1 \ {\rm at} \ \Omega \to \infty.
\end{equation}
Using Eqs.~(\ref{h-av_t}), (\ref{h_aver_low}), or (\ref{h_t-hf}) one has to ensure that $b<b_c$. Value of
$b_c$
can be extracted from the
mentioned figure at given $\Omega$. Note that the stability condition should be valid at any
$X$, therefore the value of $h_0$, which minimizes
$\Phi_{MV}(\Omega h_0^2)h_0^{-5/2}$ (at each time moment) must be
substituted in Eq.~(\ref{b_c}) to determine whether the layer is
stable or not. For most of cases, except for $3< \Omega <8$, this
means that the {\em maximum value} of $h_0$ should be used. In the
opposite case the minimum value of $\Phi_{MV}(\mu)$ ($\approx
5.5$) and again the maximum of $h_0$ can be used in order to
estimate $b_c$.

For the case of low frequency one can take the acceleration $b\Omega^2$ up to
\begin{equation} \label{b_c_smOm}
b_c \Omega^2 \approx 8.5\sqrt{C}h_0^{-5/2}
\end{equation}
in Eqs.~(\ref{h_aver_low}).
In the opposite limiting case, $\Omega \gg 1$, the parameter $b\Omega$, which enters Eq.~(\ref{h_t-hf}),
should be small:
\begin{equation} \label{b_c_largeOm}
b\Omega < b_c\Omega = \sqrt{\frac{C}{\Omega}h_0^3} \ll 1.
\end{equation}

The inequality (\ref{b_c_largeOm}) holds in the limiting case $1
\ll \Omega \ll \epsilon^{-1}$, where the Faraday instability is
caused by the longwave perturbations and the dissipation is
negligible beyond the boundary layers near the rigid wall and free
surface. Similar to Eq.~(\ref{b_c_largeOm}) the stability bound
$b_c\Omega$ is small even for $\Omega=O(\epsilon^{-1})$; the
perturbations with moderate wavelength are critical in this case.
Thus, in this case $b\Omega$ must also be small to prevent the
parametric instability.

For very large frequency, i.e. $\Omega \gg \epsilon^{-1}$, the
volume dissipation prevails, which leads to another limitation on
the amplitude. Indeed, it follows from Ref. \cite{Mancebo} that
the inequality
\begin{equation}
\tilde b \sqrt{\frac{\omega}{\nu}} <
A_{MV}\left(\frac{\omega\rho^2\nu^3}{\sigma^2}\right)
\end{equation}
(in dimensional form) or
\begin{equation}
b \sqrt{\Omega} < \epsilon
A_{MV}\left(\Omega\epsilon^4C^{-2}\right)
\end{equation}
(in dimensionless form) must hold true to prevent the Faraday
instability. Here $A_{MV}(\mu)$ is the function obtained by
Mancebo and Vega \cite{Mancebo}, which has the following
asymptotics:
\begin{equation}
A_{MV}\left(\mu\right)\approx \left(256\mu\right)^{1/6} \ {\rm at}
\ \mu \ll 1.
\end{equation}
This gives the following critical velocity:
\begin{equation}
b_c\sqrt{\Omega}
=\epsilon\left(\frac{256\Omega\epsilon^4}{C^2}\right)^{1/6}=\left(\frac{16\sqrt{\Omega}\epsilon^5}{C}\right)^{1/3}.
\end{equation}
Multiplying this relation by $\sqrt{\Omega}$ we obtain:
\begin{equation}
b_c\Omega =\left(\frac{16\Omega^2\epsilon^5}{C}\right)^{1/3},
\end{equation}
i.e. the vibration is non-negligible in Eq.~(\ref{h_t-hf}) at high
frequency only if $\Omega^2\epsilon^5$ is finite or large. This
provides, of course, the usual limitation on the amplitude and the
frequency of the vibration in the case of inviscid pulsatile
motion (see also Refs.~\cite{LChbook,AA,Lapuerta}).

It must be emphasized that advective terms, being proportional to $u_0$, are not important for short waves,
since the advective term $ik u_0 u$ remains small in comparison with $u_t$
because of the dispersion relation $\omega^2=Cak^3$ at large $k$.

\subsection{Analysis of the nondeformable mode} \label{ssec:nondef}

The above stability analysis deals
with the capillary waves, which are based on the surface deviations.
If $b < b_c$, where $b_c$ is given by Eq.~(\ref{b_c}), the perturbations of this type decay.

However, there is also a mode, which corresponds to the
nondeformable surface. Indeed, it is obvious that the
finite-amplitude plane-parallel flow with the profile $u_0$
becomes unstable at certain intensity of the motion. Let us
analyze this mode in detail. Setting $\hat \xi=0$ in
Eqs.~(\ref{pert_eq_k}) and (\ref{pert_bcs_k}) gives:
\begin{subequations}
\label{ndfm_pert_eq}
\begin{eqnarray}
ik \hat u + \hat w_z &=& 0,
\\
\hat u_t+ik u_0 \hat u + \hat w u_{0z}  &=& - ik \hat p + \nabla ^2 \hat u,
\\
\hat w_t +ik u_0 \hat w &=& - \hat p_z + \nabla^2 \hat w,\\
\hat u &=& 0, \ \hat w = 0 \ \ {\rm at} \ z = 0, \\
\label{bc_fs_pert} \hat w&=&0, \ \hat u_z  =0 \ {\rm at} \ z = h_0.
\end{eqnarray}
\end{subequations}
This problem is one of stability for the periodic in time,
plane-parallel flow. In view of Eqs.~(\ref{sol_puls}) and
(\ref{stab_basic}) this problem is characterized only by the
parameters $b$, $\Omega$ and ``local'' values of $h_0$ and
$h_{0X}$. Due to well-known Squire theorem \cite{squire-33} there
is no need in analysis of 3D perturbations -- 2D one are critical.
Moreover, the base state should not necessarily be 2D in the
entire layer. It is sufficient that the flow is {\em locally} 2D
at any point in the $X-Y$ plane.

Note that according to the analysis in Appendix
\ref{app:p-flow} the problem (\ref{ndfm_pert_eq}) for the base
flow (\ref{stab_basic}), but posed on the interval $0<z<2h_0$ and
with the no-slip condition at $z=2h_0$ instead of
Eq.~(\ref{bc_fs_pert}), is the conventional stability problem of
the oscillatory Poiseuille flow (i.e., the flow which arises due
to the periodically oscillating longitudinal pressure gradient
$P_x=P_1\cos\Omega t$).
The more general problem, where the pressure gradient equals
$P_0+P_1\cos\Omega t$, has been investigated already, see
Refs.~\cite{Singer,Straatman,Davis_annu} and references therein.
Due to symmetry the latter problem can be split into problems for
``even'' and for  ``odd'' perturbations, meaning that $w$ is even
or odd function of the coordinate $z-h_0$. For the odd mode both
$w$ and $u_z$ vanish at $z=h_0$, which coincides with the boundary
condition (\ref{bc_fs_pert}). Therefore, the problem under
consideration is the particular case of the stability problem for
the oscillatory Poiseuille flow. However, to the best of our
knowledge there is no detailed analysis of the flow stability for
the particular case we need, i.e. $P_0=0$.

It is easy to consider two limiting cases, i.e. the high
frequency and the low frequency. The first case, in view of
Eq.~(\ref{high_freq_puls}) and (\ref{sol-bound}) is reduced to the
stability of a Stokes layer. The latter is known to be stable
\cite{Davis_annu,Kerczek}.

At $\Omega \to 0$ we have $u_0 \propto b\Omega^2$ from
Eqs.~(\ref{puls_complex}) and (\ref{sol_puls_lf}). On the other
hand in this case we can ``freeze'' the evolution of the flow and
assume that
\begin{equation} \label{u0_om0}
u_0=-\frac{b \Omega^2}{2} h_X z(2h_0-z).
\end{equation}
This means that we assume the frequency so low that the
perturbations either grow or decay before the flow in fixed point
of the layer will change itself.

In view of the above-mentioned symmetry properties, the problem
(\ref{ndfm_pert_eq}) with $u_0$ given by Eq.~(\ref{u0_om0}) is
identical to the stability problem for the {\em stationary}
Poiseuille flow in the entire layer ($h_0$ being the half of the
layer thickness), but only for the odd perturbations. If we
introduce the Reynolds number based on the velocity of the flow at
$z=h_0$, we obtain:
\begin{equation}
Re=\frac{b\Omega^2 h_X h_0^3}{2}.
\end{equation}
[It follows from Eq.~(\ref{ndfm_pert_eq}) that the viscosity is equal to unity.]

It is known \cite{Orszag} that the critical value of the Reynolds
number is $5772$, i.e. the flow remains stable for
\begin{equation}
Re < Re_c=5772.
\end{equation}
Thus the flow forced by low frequency vibration is stable if
\begin{equation}
b\Omega^2 <b_{c1}\Omega^2 =\frac{11542}{h_X h_0^3}.
\end{equation}
Of course, this limitation is less severe than
Eq.~(\ref{b_c_smOm}) for any reasonable value of $h_X$. The
product $h_X h_0^3$ should be maximized over the longitudinal
coordinate $X$ at each moment of time. Thus, the pulsatile flow is
stable at low frequency up to finite value of the acceleration,
i.e. the vibration can produce the finite impact in this limiting
case. The complete analysis of the problem (\ref{ndfm_pert_eq})
will be performed elsewhere. This research will provide the
threshold value of the amplitude
\begin{equation}\label{bc_Pois}
Re_c(\Omega h_0^2)=b_{c1}\Omega |I(\alpha h_0)| h_0 h_X.
\end{equation}
(We again define the Reynolds number via the velocity at the
center of the layer, $z=h_0$.) Recall that due to the minimization
of $b_{c1}$, the {\em maximal} value of $|I(\alpha h_0)| h_0 h_X$
should be used in Eq.~(\ref{bc_Pois}).

To conclude, we showed in this section that there exists an upper
bound $b=b_c$ below which the pulsatile flow is stable. The
further analysis as well as the results of
Secs.~\ref{sec:puls}-\ref{sec:3D} are based on the assumption
$b<b_c$.

\section{Evolution of the perturbations to the flat layer} \label{sec:layer}
\subsection{Linear stability analysis} \label{ssec:linear}

The amplitude equation (\ref{h-av_t}) has the obvious solution
$h_0=1$ corresponding to the equilibrium state. It follows from
Eqs.~(\ref{sol_puls}) and (\ref{aver}) that
\begin{subequations} \label{basic}
\begin{eqnarray}
h_0&=&1, \ p_0=G_0(1-Z)+\phi(1), \ q_0=\Omega(1-Z), \\
U_0&=&W_0=I_0=K_0=H_0=0.
\end{eqnarray}
\end{subequations}
Thus in the reference frame of the substrate the fluid is
motionless -- vibration only adds the oscillatory component to the
pressure field.

Let in Eq.~(\ref{h-av_t}) $h = 1+\xi$, where $\xi$ is small
perturbation. Linearizing with respect to $\xi$ we obtain:
\begin{equation} \label{xi_t_aver}
\xi_T=\frac{1}{3}\left\{ \left[\phi'(1)+
G_0\right]\xi-\left[C+\frac{b^2 \Omega^2 f_r(\alpha)}{2}\right]
\xi_{XX} \right\}_{XX},
\end{equation}
where $\alpha^2=-i\Omega$ and $f(y)$ is given by
Eq.~(\ref{f_definition}).

The {\em linear} stability of the layer without accounting
for van der Waals attraction, and for the opposite direction of
the gravity field
 was studied by Lapuerta
et al. \cite{Lapuerta}. In this setup the Rayleigh-Taylor
instability emerges. The equations governing the dynamics of
small perturbations derived in Ref.~\cite{Lapuerta}  [see
Eqs.~(2.35) there] can be obtained from Eq.~(\ref{xi_t_aver}).

However, even in this case there is an important difference in the
interpretation of the results. Lapuerta et al. \cite{Lapuerta}
address the case of {\em finite} vibration amplitude $B$ and
consequently, they find that the influence of the vibration is
{\em small}. To gain a finite impact of the vibration they proceed
to the detailed analysis in the limit of high vibration
frequencies, $\Omega \gg 1$.  Conversely, in this paper {\em
large} vibration amplitude is considered and we show that the
Eq.~(\ref{xi_t_aver}) remains valid even in this case. Thus, we
extend the domain of applicability of the results obtained by
Lapuerta et al. by showing that the {\em finite impact} of the
vibration is possible even at moderate frequencies, which is
most important for thin films.

The typical stability curves are shown in Ref. \cite{Lapuerta}:
Figure~4 there presents the dependence of the dimensionless
amplitude of the vibration $A_L$ on the dimensionless frequency
$\omega_L$ for different values of the parameter $\alpha_L$, which
is proportional to the surface tension. The results of our linear
stability analysis (for $\phi=0$) can be extracted from this
figure when the following substitutions are made:
\begin{equation}
A_L=k\frac{b\Omega^2}{3A-G_0}, \ \omega_L=\Omega, \
\alpha_L=\frac{k^2C}{3A-G_0}.
\end{equation}
[Recall that $\phi(h)=Ah^{-3}$.] However, to avoid the
recalculation we present the results of stability analysis below.

Seeking the solution in the form of a plane wave
$\xi=\xi_0\exp\left(-\lambda T+ikX\right)$, where $\lambda$ is the
decay rate and $k$ is the wavenumber, we obtain
\begin{equation}\label{decay}
\lambda=\frac{1}{3}k^2\left\{ \phi^{\prime}(1)
+G_0+k^2\left[C+\frac{1}{2}b^2 \Omega^2
f_r(\alpha)\right]\right\}.
\end{equation}
The stability criteria ($\lambda=0$) is
\begin{equation} \label{stab0}
\phi'(1)+ G_0 + \left[C+\frac{1}{2}b^2 \Omega^2f_r(\alpha)\right]
k^2=0.
\end{equation}
Thus the vibration and  the surface tension do not damp the
longwave instability: the perturbations with small
$k$ grow at $\phi'(1)+G_0<0$. However, in confined
cavities with large aspect ratios the spectrum of the wavenumbers
is discrete and bounded from below. Therefore the impact of the
vibration and the surface tension becomes determinative in this
case.

From Eq.~(\ref{stab0}) one can see that the critical value of the
wavenumber, $k_c$ (i.e the value that corresponds to vanishing
growth rate of the perturbation) becomes smaller due to vibration:
\begin{equation} \label{kc}
k_c^2=-\frac{\phi'(1)+ G_0} {C+\frac{1}{2}b^2
\Omega^2f_r(\alpha)}.
\end{equation}
Again we note that $f_r$ grows monotonically from zero (at
$\Omega\to 0$) to unity (at $\Omega \to \infty$), see
Fig.~\ref{fig:Qf}(a). Thus the vibration leads to the {\it
stabilization} of the thin film. This stabilization effect is
obviously augmented with the increase of the frequency, even when
this increase is accompanied by the decrease of $b$ to keep fixed
the power of the vibration, $b^2\Omega^2$.

At large $\Omega$ Eq.~(\ref{stab0}) reduces to
\begin{equation} \label{stab_hf}
\phi'(1)+ G_0 + \left(C+\frac{b^2 \Omega^2}{2}\right) k^2=0.
\end{equation}
A similar equation (without the first term and with negative
$G_0$) is the well-known result on the suppression of the
Rayleigh-Taylor instability \cite{LChbook,AA,Lapuerta}.

\begin{figure*}[!t]
\includegraphics[width=12.0cm]{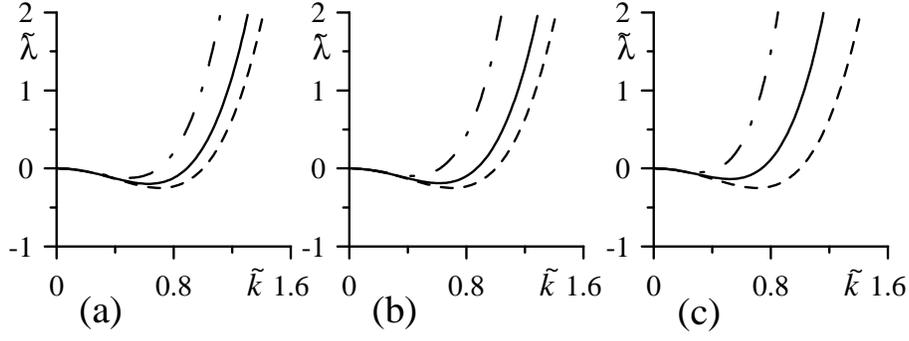}
\caption{Linear decay rate of the perturbation, $\tilde \lambda$
[Eq.~(\ref{decay1})] as function of the wavenumber $\tilde k$.
$\tilde G_0 =3.33\cdot 10^{-4}$. (a): $\Omega=0.2, \ V=0, \, 50,
\, 200$ (dashed, solid, and dash-dotted lines, respectively); (b):
$\Omega=2$; (c): $\Omega=20$. For (b) and (c) dashed, solid, and
dash-dotted lines correspond to $\ V=0, \, 1, \, 5$, respectively.
} \label{fig:decay}
\end{figure*}
\begin{figure}[!t]
\includegraphics[width=8.0cm]{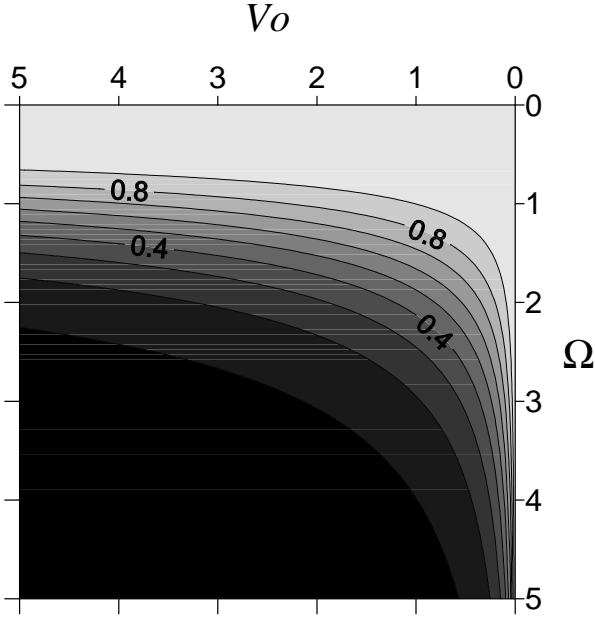}
\caption{Critical value of the wavenumber $\tilde k_c$ as function
of $Vo\equiv V\Omega^{-2}$ and the frequency $\Omega$. $\tilde G_0
=3.33\cdot 10^{-4}$.} \label{fig:kc}
\end{figure}

Introduction of the rescaled wavenumber $\tilde k=k\sqrt{C/3A}$
and the growth rate $\tilde\lambda=\lambda C/\left(3A^2\right) $
results in the following expressions for $\tilde\lambda$:
\begin{equation}\label{decay1}
\tilde \lambda=\tilde k^2\left\{\tilde G_0-1 +\tilde k^2\left[1+V
f_r(\alpha)\right]\right\},
\end{equation}
and for the critical wavenumber:
\begin{equation} \label{kc_resc}
\tilde k_c^2=\frac{1- \tilde G_0} {1+V f_r(\alpha)}.
\end{equation}
Here
\begin{equation}\label{resc_par}
\tilde G_0= \frac{G_0}{3A}, \ V=\frac{b^2 \Omega^2}{2C}=\frac{B^2
\Omega^2}{2Ca}
\end{equation}
are the rescaled Galileo number and the vibrational parameter,
respectively. The Galileo number usually is quite small for thin
films. We take $\tilde G_0=3.3\cdot 10^{-4}$. This corresponds to
$A^{\prime}=6\pi\cdot 10^{-21}J$ (value for water) and $\hat
H_0=1000 \AA$. Dependence of $\tilde \lambda(\tilde k)$ is shown
in Fig.~\ref{fig:decay} for various values of $\Omega$ and $V$.
Clearly, $\tilde k_c$ decreases with $V$. Moreover, as it is shown
in Fig.~\ref{fig:kc}, $k_c$ also decreases with growth
 of $\Omega$. Thus the vibration stabilizes the film.
And, stabilization is more pronounced for larger values of
$\Omega$.

\subsection{Weakly nonlinear analysis}\label{ssec:weakly}

Let consider the behavior of perturbations near the stability
threshold determined in the previous subsection. For this purpose
it is convenient to rescale the time and the coordinate as
follows:
\begin{equation}
\tilde X = \sqrt{\frac{3A}{C}} X, \ \tilde T= \frac{3A^2}{C} T.
\end{equation}
Substituting these relations into Eqs.~(\ref{h-av_t}) one can
obtain
\begin{subequations}\label{h-av_t-resc}
\begin{eqnarray}
h_{\tilde T}&=&\left(h^3 \tilde\Pi_{\tilde X}-3V Q\right)_{\tilde X}, \\
\tilde \Pi& \equiv& \frac{1}{3h^3}- h_{\tilde X \tilde X} + \tilde
G_0 h
+V{\rm Re} H,\\
Q&=&Q_1(\gamma)h^2h_{\tilde X}^3+Q_2(\gamma)h^3h_{\tilde X}
h_{\tilde X \tilde X},
\end{eqnarray}
\end{subequations}
where $Q_{1,2}$ are given by Eqs.~(\ref{coeff}) and the rescaled
parameters defined by Eq.~(\ref{resc_par}) are used. Below we use
only the rescaled coordinate and the time. Thus we omit the tildes
above $X$ and $T$.

In Sec.~\ref{ssec:linear} it has been shown that the growth rate
is real at the stability threshold. Consequently, the branching
solution is stationary, i.e. one can omit the left-hand side of
the amplitude equation to study direction of branching only. This
also allows us to integrate Eq.~(\ref{h-av_t-resc}) once. Besides,
seeking the solution with fixed wavenumber $\tilde k$ we introduce
the variable $\zeta=\tilde k X$. The surface deflection $h$ now is
a $2\pi$-periodic function of $\zeta$, and it solves the following
equation:
\begin{equation}\label{h-av_stat}
h^3 \left(\frac{1}{3h^3}- \tilde k^2 h_{\zeta\zeta} + \tilde G_0 h
+V \tilde k^2{\rm Re} H\right)_\zeta-3V \tilde k^2 Q=const.
\end{equation}
Note that $X$ must be replaced with $\zeta$ in the expressions for
$Q_{1,2}$ and $H$.

Next, we expand $h$ near the base solution
(\ref{basic}) as follows:
\begin{equation}
h=1+\delta h_1 + \delta^2 h_2 + \ldots.
\end{equation}
The wavenumber is assumed close to the critical value $k_0$,
i.e.
\begin{equation}
\tilde k=\tilde k_0+\delta^2 \tilde k_2 + \ldots.
\end{equation}
Substituting these expansions into Eq.~(\ref{h-av_stat}), we arrive
at order zero to
\begin{equation}
\hat L h_1\equiv (1-\tilde G_0)h_1^\prime+\tilde k_0^2\left[1+V
f_r(\alpha)\right]h_1^{\prime\prime\prime}=0,
\end{equation}
where the prime denotes the derivative with respect to $\zeta$.
Solution of this equation is
\begin{equation}
h_1=a \cos \zeta.
\end{equation}
The wavenumber is given by the expression
\begin{equation}
\tilde k_0^2=\frac{1-\tilde G_0}{1+V f_r(\alpha)},
\end{equation}
which obviously coincides with Eq.~(\ref{kc_resc}). Recall that we
are interested in the case $\tilde G_0 < 1$ which holds true for
the thin film.

The equation at the second order in $\delta$ has the form:
\begin{eqnarray}
\nonumber \hat L h_2&=&2(h_1^2)^\prime+V \tilde k_0^2F_1 \left(h_1
h_1^{\prime\prime\prime}+3h_1^\prime h_1^{\prime\prime}\right)
\\
&& -3V\tilde k_0^2Q_2h_1^\prime h_1^{\prime\prime}.
\end{eqnarray}
The solution of this equation is
\begin{eqnarray}
h_2&=&C_1 a^2 \cos 2\zeta, \\ \nonumber C_1&=&-\frac{1}{3(1-\tilde
G_0)}\left[1+V \tilde k_0^2\left(F_1+\frac{3}{4}Q_2\right)\right].
\end{eqnarray}
Hereafter
\begin{equation}
F_1=\left(\frac{dF}{dh}\right)_{h=1}, \
F_2=\left(\frac{d^2F}{dh^2}\right)_{h=1}, \ F\equiv hf_r(\alpha
h).
\end{equation}
At the third order
we need only the solvability condition. (We do not present here the corresponding equation for $h_3$.)
This condition couples the
correction to the wavenumber $k_2$ to the amplitude of
the perturbation $a$, as follows:
\begin{widetext}
\begin{equation}
\tilde k_2=\frac{\tilde k_0a^2}{2\left(1-\tilde
G_0\right)}\left[\frac{5}{2}-2C_1+V\tilde
k_0^2\left(\frac{9}{4}Q_1+3Q_2C_1-\frac{3}{4}Q_2^\prime-\frac{1}{2}C_1F_1-\frac{1}{8}F_2\right)\right].
\end{equation}
\end{widetext}
Our numerical simulations show that the expression in the square
brackets is always positive. Thus $\tilde k_2>0$, i.e. a small
amplitude solution exists at $\tilde k>\tilde k_0$. This means that the
subcritical bifurcation takes place.

To summarize, the solution emerging at $\tilde k = \tilde k_0$ is
unstable and a finite amplitude excitation (probably leading to
rupture) is expected.

\subsection{Nonlinear evolution of perturbations}

In this Section we analyze the finite-amplitude deflections of the
free surface, starting from stationary solutions $h_s$ of the
amplitude equation. For this purpose we look for the periodic
(with respect
 to $\zeta$) solutions of Eq.~(\ref{h-av_stat}).

Due to evident symmetry properties of the solution one can
integrate Eq.~(\ref{h-av_stat}) over half of the period, vanishing
all the odd derivatives at $\zeta=0, \pi$. This means that
$const.=0$ in Eq.~(\ref{h-av_stat}). We also have two boundary
conditions:
\begin{equation} \label{bc_zeta}
h_s^{\prime}=0 \ {\rm at} \ \zeta=0,\pi.
\end{equation}
Besides, the stationary solution conserves the liquid volume:
\begin{equation}
\int_0^\pi (h_s -1)d\zeta=0.
\end{equation}
This provides the third boundary condition for the third order
ODE, Eq.~(\ref{h-av_stat}), completing the problem statement.

\begin{figure*}[!t]
\includegraphics[width=12.0cm]{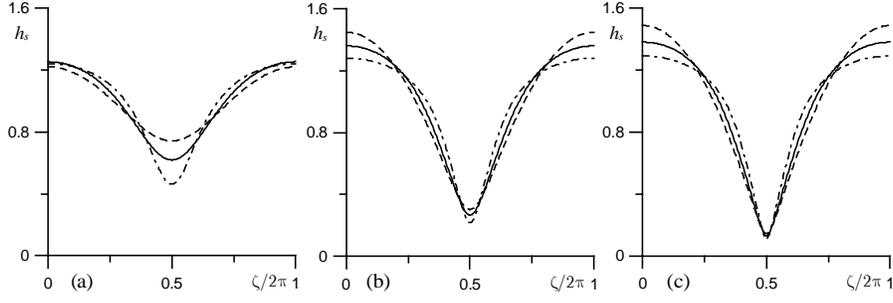}
\caption{Unstable stationary solutions computed from
Eq.~(\ref{h-av_stat}) for $\tilde G_0=3.33\cdot 10^{-4}, \
\Omega=0.2$. (a): $\tilde k=1.1$, (b): $\tilde k=2.2$, (c):
$\tilde k=4.4$. Dashed, solid, and dash-dotted lines correspond to
$V=0, \, 50, \, 200$.} \label{fig:shape_stat}
\end{figure*}
\begin{figure*}[!t]
\includegraphics[width=12.0cm]{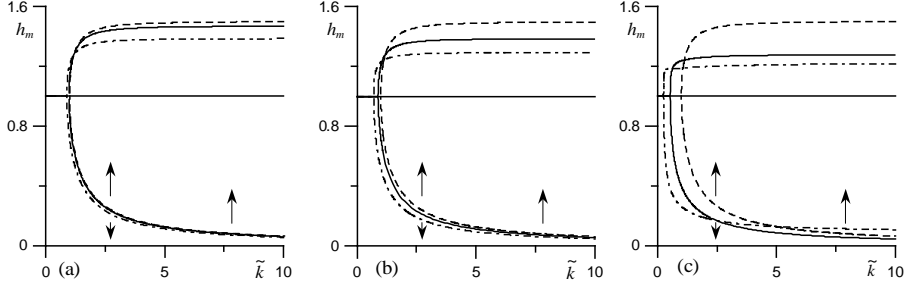}
\caption{Minimal (lower branch) and maximal (upper  branch)
thickness of the film at the corresponding unstable stationary
solution. $\tilde G_0=3.33\cdot 10^{-4}$. (a): $\Omega=0.02$,
dashed, solid, and dash-dotted lines correspond to $V=0, \, 800,
\, 5000$ ($\tilde k_0=0.99998, \, 0.9793, \, 0.8885$); (b):
$\Omega=0.2$, dashed, solid, and dash-dotted lines correspond to
$V=0, \, 50, \, 200$ ($\tilde k_0=0.99998, \, 0.8891, \, 0.6968$);
(c): $\Omega=2$, dashed, solid, and dash-dotted lines correspond
to $V=0, \, 8, \, 50$ ($\tilde k_0=0.99998, \, 0.5287, \,
0.2418$). Arrows show boundaries of domains of attraction. }
\label{fig:h_stat}
\end{figure*}

\begin{figure}[!t]
\includegraphics[width=5.0cm]{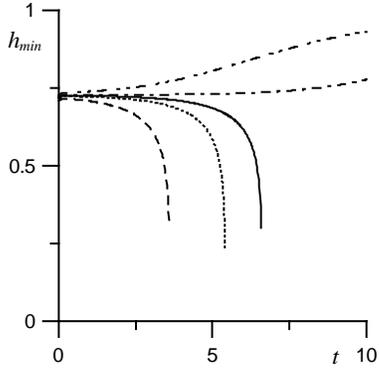}
\caption{Evolution of the minimum film thickness, as given by
direct numerical computation of Eq.~(\ref{h-av_t}). Unstable
stationary solutions with small perturbation of the free surface
$\delta_h$ were chosen for the initial condition. $\tilde
G_0=3.33\cdot 10^{-4}, \ \tilde k = 1.1, \ \Omega=0.02, \ V =
800$. $\delta_h=0.009, \, 0.002, \, 0, \, -0.002, \, -0.01$
(dash-double-dotted, dash-dotted, solid, dotted, and dashed lines,
respectively). } \label{fig:evol}
\end{figure}

The shooting method is applied to numerically integrate this
boundary value problem. Some results are presented in
Figs.~\ref{fig:shape_stat} and \ref{fig:h_stat}. It is clearly
seen that only the subcritical (and, consequently, unstable)
solution branch is present, i.e. there is no bifurcation except
the inverse pitchfork bifurcation studied in
Sec.~\ref{ssec:weakly}. Therefore, the lower branches in
Fig.~\ref{fig:h_stat} are the boundaries of domains of attraction:
the initial perturbation with $min[h(\zeta, T=0)]>h_m$ decays and
the free surface becomes flat, while in the opposite case the free
surface is attracted to the solid, which leads to rupture. These
predictions are in good agreement with the results of the
numerical simulation of Eqs.~(\ref{h-av_t-resc}), as shown in
Fig.~\ref{fig:evol}.

The upper branches in Figs.~\ref{fig:h_stat} are of importance for
the stability of the oscillatory flow (see Sec.~\ref{sec:stab}).
Indeed, Eq.~(\ref{b_c}) requires the {\em maximal} value of the surface
deviation $h_0$, which can be extracted from
Fig.~\ref{fig:h_stat}.

It is important to point out that the increase of the vibration
amplitude amplifies the film stability: at fixed $\tilde k$ and
with $\Omega$ increasing, the initial deviation of the flat
surface decays at larger values of $V$. The only exception is
provided by $\Omega=2$. In this case $h_m$ is nonmonotonic
function of $V$ starting from certain value of $\tilde k$. Also it
is interesting that the maximal deviation of the surface decreases
with growth of $V$: the surface tends to become flat as it is
shown in Fig.~\ref{fig:shape_stat}. In some sense this tendency is
reminiscent of the known averaged behavior \cite{LChbook,LChMcgr}:
the free surface/interface tries to orient normally to the
vibration axis. But it is necessary to keep in mind that the
surfaces shown in Fig.~\ref{fig:shape_stat} correspond to {\em
unstable states}.

\section{Summary}\label{sec:conc}

In this paper the impacts of the vertical vibration on the
dynamics of the thin liquid film are analyzed. The set of
equations governing the averaged dynamics of the fluid flow and
the nonlinear, fourth-order amplitude equation (\ref{h-av_t}) [or
(\ref{h-av_t_3D}) in 3D case] describing the averaged evolution of
the film thickness are obtained in the lubrication approximation.

We use the paradoxical  scaling (at least at the first glance),
assuming (i) the vibration period is
{\em comparable} to the characteristic time of the momentum
relaxation across the layer and (ii) the vibration amplitude is
{\em large} in comparison with the mean layer
thickness. The first condition (termed ``moderate frequency"
or ``finite frequency") allows us to consider ultrathin liquid
layers within the framework of the averaging method. The second
condition warrants that the impact of the vibration is not
vanishingly small. Using the results from Refs.
\cite{Mancebo,Straatman,Singer} we prove that these assumptions do
not necessarily lead to the parametric instability.

Indeed, we show that the stability problem for the pulsatile flow
can be separated into the problems for the deformable and
non-deformable modes. The former problem reduces to the analysis
of Faraday instability, while the latter problem reduces to the
stability analysis of the oscillatory Poiseuille flow. Analyzing
well known results obtained for these two problems we deduce that
there exists a window of stability: below certain threshold
intensity of the vibration [Eqs.~(\ref{b_c}) and
(\ref{bc_Pois})] the pulsatile flow is stable, while the averaged
effects are well pronounced.

The analyses of the averaged dynamics of the thin film demonstrate
the strong stabilizing impact of the vibration. First, the
vibration damps the short-wavelength instability. In other words,
it decreases the critical cut-off wavenumber $k_c$, such that
instability occurs at $k<k_c$ only [see Eqs. (\ref{decay}),
(\ref{kc}) and Figs.~\ref{fig:decay} and ~\ref{fig:kc}]. In this
sense the vibration acts in a way similar to the surface tension.
Therefore, in order to prevent a longwave instability one can use
a cavity of horizontal size $L <L_c \sim k_c^{-1}$, which is
larger in presence of the vibration. Second, the vibration
augments the domain of attraction of the flat undeformed surface,
i.e. larger initial surface deflections decay (see
Fig.~\ref{fig:h_stat}) or, in other words, larger initial
distortions of a flat surface are admissible without occurrence of dewetting.

Thus, the vertical vibration of moderate frequency is the
effective method of control of the thin film instability. This is
especially important since the standard (high-frequency)
approximation cannot be applied to thin films.

\section{Acknowledgments}

S.~S. was partially supported by the Fund ``Perm Hydrodynamics''.
A.~A.~A. acknowledges CRDF for the financial support within the
framework of grant Y3-MP-09-01.

\appendix

\section{Reduction of pulsatile flow to the oscillatory Poiseuille flow}
\label{app:p-flow}

In this section we show that the pulsatile flow
given by Eqs.~(\ref{puls_complex}) and (\ref{sol_puls}) can be reduced to the well-
known oscillatory Poiseuille flow. Such a transformation is useful
for the stability analysis carried out in
Sec.~\ref{ssec:nondef}. Throughout this section we omit the tildes
over the velocity and the pressure field, with the understanding that only the
oscillatory components are involved.

In order to obtain the oscillatory Poiseuille flow we return
to the unscaled coordinates $x$ and $z$, setting {\em locally}
\begin{equation}\label{hx_const}
h=h_0+O(\epsilon), \ h_X=const + O(\epsilon).
\end{equation}
These equations mean that the layer thickness changes slowly with
$x$. Thus one can assume the constant thickness $h_0$. In view of the
scaling (\ref{pres_scal}) and Eqs.~(\ref{puls_complex}) and
(\ref{sol_puls}) the longitudinal pressure gradient is $b \Omega
h_X\cos \tau$. Thus the pressure gradient is spatially
uniform but oscillates in time. This means that some kind of the so-called oscillatory
Poiseuille flow is under consideration.

Neglecting $h_{XX}$ according  to Eq.~(\ref{hx_const}) we
arrive at the following expressions for the amplitudes of
pulsations
\begin{subequations}\label{sol_hx}
\begin{eqnarray}
I&=&i h_X\left(1-\frac{\cos\alpha (h_0-z)}{\cos\alpha h_0}\right), \\
K&=&i
 h_X^2\frac{1-\cos\alpha
Z}{\cos^2\alpha h_0}, \ H=h_X^2\tan^2\alpha h_0,
\end{eqnarray}
\end{subequations}
which along with the scalings (\ref{pres_scal}) and
(\ref{rescaled_velo}) gives:
%
\begin{eqnarray}
\label{sol_hx_t}u&=&b \Omega h_X {\rm Re} \left[i \left(1-\frac{\cos\alpha (h_0-z)}{\cos\alpha h_0}\right)e^{i\tau}\right], \\
\label{wh} w&=&O(\epsilon),  \ h-h_0=O(\epsilon).
\end{eqnarray}
%
Thus this case corresponds to oscillatory 1D flow in a {\em
locally} flat layer under the spatially uniform longitudinal
gradient of pressure. Note that the second relation in Eqs.
(\ref{wh}) justifies the first assumption in
Eqs.~(\ref{hx_const}).

Separating the real and imaginary parts in Eq.~(\ref{sol_hx_t}) we
arrive at the following expressions for the $x$-component of the
pulsation velocity:
\begin{widetext}
\begin{eqnarray}\label{puls_real}
u &=& -b \Omega h_X
\left\{\left[1-\frac{a_S(h_0)a_S(z)+b_S(h_0)b_S(z)}{a_S^2(h_0)+b_S^2(h_0)}
\right]\sin\tau
-\frac{a_S(h_0)b_S(z)-a_S(z)b_S(h_0)} {a_S^2(h_0)+b_S^2(h_0)}\cos
\tau\right\},\\
\nonumber a_S(z)&\equiv &\cos\alpha_rz\cosh\alpha_rz, \
b_S(z)\equiv \sin\alpha_rz\sinh\alpha_rz, \
\alpha_r=\sqrt{\Omega/2}.
\end{eqnarray}
\end{widetext}

Such flow, but in a gap between two rigid boundaries, is well
studied -- see, for instance, the recent papers by Singer et al.
\cite{Singer} and Straatman et al. \cite{Straatman}, who address
the stability problem for the flow. (Usually the modulated
Poiseuille flow is considered, i.e. the pressure gradient
oscillates about a non-zero mean value.)

Due to symmetry the oscillatory Poiseuille flows in a layer with
the {\it free nondeformable} surface at $z=h_0$ and in a layer
with the upper {\it rigid} wall at $z=2h_0$ are
identical. Indeed, in the latter case in
the plane of symmetry $z=h_0$ the ``no-stress'' condition $u_z=0$
is obviously held.

Thus, introducing variable $y_S= z-h_0$ in Eq.~(\ref{puls_real})
we obtain the velocity profile of the oscillatory Poiseuille flow
[cf. oscillatory part of Eq.~(2.2) in Ref.~\cite{Singer}].

\section{Solution of the set of averaged equations}
\label{app-aver}

Solution of the problem for the averaged fields can be represented
as the sum of two solutions. The first one is conventional (see,
for example, Ref.~\cite{Oron}); we mark it by the subscript
``$c$'':
\begin{subequations}\label{aver-sol1}
\begin{eqnarray}
\nonumber p_c &\equiv & \Pi-G_0Z\\
&=&\phi-C h_{XX} +G_0(h-Z) +
\frac{b^2\Omega^2}{2}{\rm Re}H, \\
\label{Uc} U_c&=&\frac{1}{2}Z(Z-2h)\Pi_X, \\
W_c&=&-\frac{1}{3}\left[Z^2(Z-3h)\Pi_X\right]_X.
\end{eqnarray}
\end{subequations}
Note that this solution leads to the first term in the evolution
equation for the film thickness, Eq.~(\ref{h-av_t}). This part of
the solution coincides with Eqs.~(2.47) in Ref.~\cite{Oron} up to
the averaged correction to the pressure field.

The second part is the solution of the nonhomogeneous boundary value
problem with the remaining vibration-generated terms at the
right-hand sides. Using subscript ``$v$'' for this part of
solution, we rewrite it in the following form:
\begin{equation}
\label{aver-sol2} p_v=0, \ U_v=\frac{1}{2}b^2\Omega^2 U^{(v)}, \
W_v=\frac{1}{2}b^2\Omega^2W^{(v)},
\end{equation}
where $U^{(v)}$ and $W^{(v)}$ solve the following boundary value
problem:
\begin{subequations}
\begin{eqnarray}\label{aver-v}
U^{(v)}_{ZZ}&=&{\rm Re} \left(I^* I_X+ K^*I_Z\right), \ W^{(v)}_Z=-U_X^{(v)}, \\
U^{(v)}&=&W^{(v)}=0 \ {\rm at} \ Z = 0, \\
U_Z^{(v)}&=&-\Omega^{-1} {\rm Re} \left(I_{ZZ}^* H\right) \ {\rm
at} \ Z = h.
\end{eqnarray}
\end{subequations}
This set of equations can be easily integrated. First,
\begin{equation}
U_Z^{(v)}=-\Omega^{-1}{\rm Re} \left[I_{ZZ}^*(h)
H\right]-\int_Z^h{\rm Re} \left(I^* I_X+ K^*I_Z\right) dZ.
\end{equation}
Accounting for Eq.~(\ref{puls}) of the pulsatile motion and
integrating by parts, one can rewrite the last expression as
\begin{equation}
U_Z^{(v)}=-h_X{\rm Re} H + {\rm Re} \left(I^*K\right) -\partial_X
\int_Z^h |I|^2 dZ.
\end{equation}
After one more integration we arrive at the following solution:
\begin{eqnarray}
\nonumber U^{(v)}&=&-Z h_X{\rm Re} H + \int_0^Z{\rm
Re}\left(I^*K\right)dZ
\\&&-\partial_X \int_0^Z d\zeta \int_\zeta^h |I(X,\xi)|^2 d \xi.
\end{eqnarray}
Evaluation of these integrals leads to the cumbersome
formulas, which we do not present here.

We also do not present the expression for $W^{(v)}$, as it is
not needed in order to obtain the evolution equation for film
thickness $h$. Indeed, this part of the solution results in
the term
\begin{equation}
-\partial_X\int_0^h U^{(v)}dZ-\Omega^{-1}{\rm Re}\{I^*H\}
\end{equation}
at the right-hand side of such an equation. This term translates to the term
$Q(h)$ in Eq.~(\ref{h-av_t}).

\end{document}